\documentclass[AMA,STIX2COL]{WileyNJD-v2}
\articletype{Article type}%
\received{XX.XX.XXXX}
\revised{YY.YY.YYYY}
\accepted{ZZ.ZZ.ZZZZ}

\usepackage{verbatim}

\providecommand{\differential}{\mathrm{d}}
\providecommand{\ks}{k_\mathrm{S}}
\providecommand{\kt}{k_\mathrm{T}}
\providecommand{\mus}{\mu_\mathrm{S}}
\providecommand{\mut}{\mu_\mathrm{T}}
\providecommand{\taus}{\tau_\mathrm{S}}
\providecommand{\taut}{\tau_\mathrm{T}}

\begin{document}

\title{Model averaging for robust extrapolation in evidence synthesis}

\author[1]{Christian R\"{o}ver}
\author[2]{Simon Wandel}
\author[1]{Tim Friede}
\authormark{C.~R\"{O}VER, S.~WANDEL, T.~FRIEDE}

\address[1]{\orgdiv{Department of Medical Statistics}, \orgname{University Medical Center G\"{o}ttingen}, \orgaddress{\state{G\"{o}ttingen}, \country{Germany}}}
\address[2]{\orgname{Novartis Pharma AG}, \orgaddress{\state{Basel}, \country{Switzerland}}}
\corres{*Christian R\"{o}ver, \email{christian.roever@med.uni-goettingen.de}}

\abstract[Summary]{ 
  Extrapolation from a source to a target, e.g., from adults to
  children, is a promising approach to utilizing external information
  when data are sparse.  In the context of meta-analysis, one is
  commonly faced with a small number of studies, while potentially
  relevant additional information may also be available. Here we
  describe a simple extrapolation strategy using heavy-tailed mixture
  priors for effect estimation in meta-analysis, which effectively
  results in a model-averaging technique.  The described method is
  robust in the sense that a potential prior-data conflict, i.e., a
  discrepancy between source and target data, is explicitly
  anticipated.  The aim of this paper is to develop a solution for this
  particular application, to showcase the ease of implementation by
  providing \textsf{R}~code, and to demonstrate the robustness of the 
  general approach in simulations.  
}


\jnlcitation{\cname{\author{C. R\"{o}ver}, 
                    \author{S. Wandel}, and
                    \author{T. Friede}} 
             (\cyear{2017}), 
             \ctitle{Model averaging for robust extrapolation in evidence synthesis}, 
             \cjournal{Statistics in Medicine}, \cvol{2018;00:0--0}.}

\maketitle

\footnotetext{\textbf{Abbreviations:} AR, acute rejection; CI,
  credible interval; MAC, meta-analytic-combined; MAP,
  meta-analytic-predictive; NNHM, normal-normal hierarchical model;
  OR, odds ratio}

\section{Introduction}
  When empirical evidence is sparse, it may be useful to be able to
  utilize related source data to extrapolate to the targeted
  population.  This is especially relevant in the context of rare
  diseases or small populations, but the problem is common in many
  applications.  Several regulatory guidelines touch upon the problem
  from different angles, either explicitly concerning extrapolation
  \cite{EMA2016} and the use of external data \cite{FDA2017}, or, for
  example, in the contexts of small populations research
  \cite{EMEA2006}, paediatric studies \cite{FDA2016}, or bridging
  studies \cite{ICH1998}.  The potential benefits of extrapolation
  approaches are generally recognized \cite{DunneEtAl2011,IRDiRC2016},
  especially in the context of rare diseases
  \cite{Dunoyer2011,HampsonEtAl2014}.  Concerning the methodological
  aspect, the use of Bayesian methods has frequently been suggested
  \cite{EMEA2006,NeuenschwanderEtAl2010,GerssKoepcke2010,FDA2016,NeuenschwanderEtAl2016b}
  and also in practice appears to be the predominant approach
  \cite{WadsworthHampsonJaki2016}.

  In a Bayesian model, external evidence may be considered e.g. via
  the formulation of informative priors or the use of hierarchical
  models
  \cite{FDA2016,WadsworthHampsonJaki2016,SchoenfeldZhengFinkelstein2009,GamaloSiebersEtAl2017}.
  As the term \emph{extrapolation} suggests, there is usually some
  doubt whether or to what extent the external data can or should be
  taken at face value and are directly applicable to the given
  context. Consequently, the implementation of (potential)
  downweighting of the external evidence is a common requirement
  \cite{WadsworthHampsonJaki2016,HlavinEtAl2016}.

  So far, the literature very much focussed on the setting of a single
  target study, but extrapolation may also be useful in evidence
  synthesis. Meta-analyses are commonly based on only few studies,
  especially when they are concerned with rare diseases
  \cite{FriedeRoeverWandelNeuenschwander2017a,FriedeRoeverWandelNeuenschwander2017b},
  but also quite generally
  \cite{DaveyEtAl2011,TurnerEtAl2012,BenderEtAl2018}.  Additional
  external data may here easily be utilized by the formulation of an
  informative prior distribution
  \cite{NeuenschwanderEtAl2010,NeuenschwanderEtAl2016b,GerssKoepcke2010}.
  In the following we introduce the implementation of some degree of
  scepticism and robustness by using a heavy-tailed mixture prior
  \cite{FuqueneCookPericchi2009,OHaganPericchi2012,SchmidliEtAl2014}.
  Computationally, a mixture prior then results in a model-averaging
  technique \citep{HoetingEtAl1999}.
  Besides the interpretation of a ``robustified'' informative prior
  distribution, this setup may also be thought of as a combination of
  several data models, corresponding to subgroupings of the data into
  studies with common or unrelated effects \cite{BornkampEtAl2016}.

  Encouraged e.g. by the
  International Rare Diseases Research Consortium (IRDiRC)
  \cite{IRDiRC2016}, the aim of this paper is to showcase the
  potential of robust extrapolation in Bayesian meta-analysis via
  prior specification as suggested by Schmidli \textit{et
    al.}\cite{SchmidliEtAl2014}, and to exemplify the relative ease of
  implementation and computation. The general idea is actually more
  widely applicable, beyond the context of meta-analysis.  We first
  decribe the general methodology in Sec.~2, and then apply the
  approach in two case studies of meta-analyses extrapolating from
  external source data (adolescents or adults) to children in
  applications in asthma and liver transplantation in Sec.~3 and~4.
  In Sec.~5, the method's long-run behaviour is investigated in a
  small simulation study.  Sec.~6 closes with some concluding remarks.
  Computations are performed using \textsf{R} and the
  \texttt{bayesmeta} and \texttt{rjags} packages
  \cite{R-Manual,bayesmeta,R:rjags}.  Example \textsf{R}~code is
  available in the Appendix.

\section{Bayesian random-effects meta-analysis}
  \subsection{The model}\label{sec:remodel}
    Meta-analyses are commonly performed using the
    \emph{normal-normal hierarchical model (NNHM)}; the model may be
    specified as follows. A number~$k$ of studies or
    measurements~$Y_i$ ($i=1,\ldots,k$) are given; these measurements
    only come with limited accuracy, as expressed by the associated
    standard errors~$s_i$. We assume that a measurement comes
    about as a draw from a normal distribution centered around the
    (study-specific) mean~$\theta_i$:
    \begin{equation}
      Y_i \,|\, \theta_i \;\sim\; \mathrm{N}(\theta_i,\, s_i^2) \mbox{.}
    \end{equation}
    The standard errors are commonly assumed known.  The
    study-specific means are not necessarily identical across studies,
    rather one allows for a certain amount of \emph{heterogeneity}
    between studies, implemented in terms of an additional variance
    component $\tau \geq 0$:
    \begin{equation}
      \theta_i \,|\, \mu,\,\tau \;\sim\; \mathrm{N}(\mu,\, \tau^2)
    \end{equation}
    \cite{HedgesOlkin,HartungKnappSinha,BorensteinEtAl,SpiegelhalterEtAl}.
    Since often primary interest lies in the overall effect~$\mu$ (and
    not in the shrinkage estimates of~$\theta_i$), the model may be
    simplified to the marginal form
    \begin{equation}
      Y_i \,|\, \mu,\, \tau \;\sim\; \mathrm{N}(\mu,\, s_i^2 + \tau^2) \mbox{.}
    \end{equation}
    There are two unknowns that one may want to infer from the data;
    the heterogeneity~$\tau$, which commonly constitutes a nuisance
    parameter, and the effect~$\mu$, which is usually of primary
    interest. In order to infer the parameters within a Bayesian
    framework, one needs to specify prior distributions for~$\mu$
    and~$\tau$.

  \subsection{Informative priors and robustness}\label{sec:infoprior}
    In the random-effects meta-analysis model we may facilitate
    extrapolation by propagating information through the analysis via
    the prior probability distribution
    \cite{NeuenschwanderEtAl2010,NeuenschwanderEtAl2016b,WadsworthHampsonJaki2016}.
    We may have information from external sources available that can
    be used to inform the analysis.  However, it is often uncertain
    whether this information is directly applicable to the present
    context or whether the possibility of an alternative model should
    also be considered \cite{HlavinEtAl2016,SchmidliEtAl2014}.  Both
    the utilization of additional information as well as the explicit
    consideration of uncertainty may also be particularly desirable in
    regulatory decision-making \citep{FDA2017,FDA2010b}.

    A simple way to implement a certain amount of scepticism is via
    two prior components, which are combined to form the prior
    distribution as a mixture
    \cite{WadsworthHampsonJaki2016,HlavinEtAl2016,HsiaoHsuTsou2007}. To
    that end we may formulate two parameter models, $M_a$ and $M_b$,
    representing the cases of equal effects for source and target data
    (where direct extrapolation would be valid), and of a different,
    unrelated effect for the new data.  The two models simply differ
    by their assumed prior for the parameters, $p(\mu,\tau|M_a)$ and
    $p(\mu,\tau|M_b)$, respectively; the associated data model and
    likelihood ($p(y|\mu,\tau)$) are identical under both models.
    Both parameter models have prior probabilities $p(M_a) \in [0,1]$
    and $p(M_b)=1-p(M_a)$ associated. The marginal prior density then
    results as
    \begin{eqnarray}\label{eqn:mixtureprior}
      p(\mu,\tau) \;=\; p(\mu,\tau|M_a) \; p(M_a) + p(\mu,\tau|M_b) \; p(M_b)
    \end{eqnarray}
    where the two components now are chosen such that
    $p(\mu,\tau|M_a)$ is informative (with probability concentrated
    according to the external information), while $p(\mu,\tau|M_b)$ is
    vague. The probability~$p(M_a)$ then reflects the certainty (or
    scepticism) associated with the external information.  The same
    approach is readily generalized to more than two components; in
    the following, we will be concerned with the case of four
    parameter models ($M_1,\ldots,M_4$) that are associated with prior
    probabilities $p(M_1),\ldots,p(M_4)$ with $\sum_i p(M_i) = 1$.

    A setup like this results in a heavy-tailed prior compared to
    $p(\mu,\tau|M_a)$ alone; such priors have favourable properties
    when the data in fact do turn out to be in conflict with the prior
    information as supplied by~$p(\mu,\tau|M_a)$ and so it will
    provide more robust inference in case of a prior/data conflict
    \cite{OHaganPericchi2012,SchmidliEtAl2014}.

  \subsection{Mixture priors and inference}\label{sec:mixprior}
    A simple mixture setup has the advantage that it simplifies
    computations; the (conditional) posteriors under the two models
    $M_a$ and $M_b$ may be computed separately, and the partial
    results then may be re-combined via their associated Bayes factors
    \cite{KassRaftery1995,SchmidliEtAl2014}. With that, the mixture
    prior effectively results in a model-averaging approach.  The
    simplicity may be seen from the derivation via Bayes'
    theorem. Consider the generic case of inferring
    parameters~$\vartheta$ from data~$y$, where the
    prior~$p(\vartheta)$ is given as a two-component mixture
    distribution analogous to~(\ref{eqn:mixtureprior}); the
    parameters' posterior distribution is given by
    \begin{eqnarray}
      p(\vartheta|y)
      &=& \label{eqn:mixtureposterior}
           p(\vartheta|y, M_a) \; p(M_a|y)  +  p(\vartheta|y, M_b) \; p(M_b|y)
    \end{eqnarray}
    (the detailed derivation is shown in the appendix).  So, from
    (\ref{eqn:mixtureposterior}) one can see that with the prior set
    up as a two-component mixture, the posterior $p(\vartheta|y)$
    again is a mixture of the two (conditional) posteriors
    ($p(\vartheta|y, M_a)$ and $p(\vartheta|y, M_b)$).  The weighting
    results from the posterior probabilities for the two model
    components ($p(M_a|y)$ and $p(M_b|y)$) which again depend on the
    marginal likelihoods ($p(y|M_a)$ and $p(y|M_b)$) and the prior
    probabilities~$p(M_i)$:
    \begin{equation}
      p(M_a|y) \;=\; \frac{p(y|M_a) \, p(M_a)}{p(y|M_a) \, p(M_a) + p(y|M_b) \, p(M_b)}
    \end{equation}
    where the marginal likelihoods are given by
    \begin{equation}
      p(y|M_i) \;=\; \int p(y|\vartheta,M_i)\, p(\vartheta|M_i) \,\differential \vartheta
    \end{equation}
    (where $i\in\{a,b\}$). The approach is analogously generalized to
    the case of more than two mixture components.

    The posterior distribution may be expressed as a
    mixture or weighted average of posterior components, placing the
    model in the class of \emph{model averaging} approaches
    \cite{HoetingEtAl1999,KassRaftery1995,Clemen1989,ClydeGeorge2004}.  
%
%
    The setup may be thought of as a combination of several
    plausible data models~$M_i$.  In this particular case, the models
    under consideration correspond to subgroups of the data into
    studies with common or unrelated parameters \cite{BornkampEtAl2016}.
    In contrast to \emph{model selection}, instead of singling out a particular model for inference, model
    averaging then allows to perform \emph{unconditional} inference by
    marginalizing over the uncertain model indicator.

    The model averaging setup considers a discrete set of potential
    data models.  The problem could alternatively be approached in
    different ways, for example, by adding further hierarchical stages
    to the model, which may then allow to encompass the same set of
    models as special cases. Conceptually, this would replace the
    ``binary'' alternatives (of exchangeable or unrelated data) by a more
    ``continuous'' notion of data similarity.

    It is important to note that, due to the dependence on marginal
    likelihoods above, the exact specification of the ``vague'' prior
    component is crucial.  The problem is related to \emph{Lindley's
      paradox} \cite{Lindley1957}: Although different vague priors may
    differ little in the posterior distributions they imply for the
    parameter vector~$\vartheta$, they may still have a substantial
    effect on the corresponding marginal likelihoods ($p(y|M_i)$), and
    with that, the eventual relative weighting of the two conditional
    posteriors via $p(\vartheta|y,M_i)$.  The effect may lead to
    somewhat counterintuitive behaviour here. While a larger prior variance
    for the vague model component may at first seem more conservative,
    it may in fact amplify the informative component's posterior
    probability by reducing the marginal likelihood under the vague
    component.

  \subsection{Meta-analysis using mixture priors}
    \subsubsection{Meta-analysis of log-ORs}
      In the following we will conduct meta-analyses using the
      random-effects model described in Sec.~\ref{sec:remodel} with a
      mixture prior as described in Sec.~\ref{sec:infoprior}
      and~\ref{sec:mixprior}.  The endpoint of interest in the
      following is a logarithmic odds ratio (log-OR).  Odds ratios and
      their standard errors on the logarithmic scale are computed
      using standard formulas \cite{HartungKnappSinha,BorensteinEtAl}.
      The mixture prior then is defined using components that imply
      different amounts or pathways of borrowing of information.

      The \emph{target} data set of primary interest consists of
      $\kt$~effect estimates $Y_{t,1},\ldots,Y_{t,\kt}$ and standard
      errors $s_{t,1},\ldots,s_{t,\kt}$.  Another set of $\ks$
      additional \emph{source} estimates and standard errors
      ($Y_{s,i}$, $s_{s,i}$) is available, which constitutes the
      potentially relevant external information.

    \subsubsection{Prior information and data pooling}
      Schmidli \textit{et~al.}\cite{SchmidliEtAl2014} pointed out that
      in the meta-analysis context consideration of external
      information may be implemented in two obvious ways. Firstly, in the
      \emph{meta-analytic-combined (MAC)} approach, both source and
      target data sets are analyzed jointly. Alternatively, one may
      perform an analysis of the source data to derive an informative
      \emph{meta-analytic-predictive (MAP)} prior distribution for the
      target data.

      Both MAC and MAP approaches can be shown to be equivalent, since
      performing separate analyses and using one result to form the
      prior for the other (MAP approach) yields identical results to a
      pooled analysis (MAC approach)\cite{SchmidliEtAl2014}.
      This has the advantage of
      making the flow of information through the analysis transparent,
      and it also may be utilized to simplify computations. Use of an
      informative prior (based on source data) may then technically
      also be viewed as a ``pooling'' of source and target data.

    \subsubsection{Vague prior}\label{sec:vagueprior}
      The vague prior ($p(\mu,\tau|M_4)$ in the following) is
      specified as uninformative, covering a range of a~priori
      plausible values.  In the following, we represent an
      uninformative prior by a zero-mean normal distribution with
      standard deviation~2 for the effect~$\mu$, and a half-normal
      distribution with scale~0.5 for the heterogeneity~$\tau$.  For
      the effect this implies a probability distribution symmetric
      around zero (corresponding to an odds ratio of~1) with a 95\%
      prior probability for the log-odds ratio $\mu$ to lie within
      $\pm 3.92$, which translates to odds ratios roughly within a
      range from~$\frac{1}{50}$ to~$50$.  This prior may be
      interpreted as equivalent to the information given by a
      contingency table of an \emph{effective sample size} of
      4~patients \citep{Roever2017}. For the heterogeneity, the
      half-normal prior also constitutes a conservative choice in the
      context of log-OR endpoints
      \cite{SpiegelhalterEtAl,FriedeRoeverWandelNeuenschwander2017a,FriedeRoeverWandelNeuenschwander2017b}.

    \subsubsection{Mixture prior components}\label{sec:datamodels}
      The informative prior components ($p(\mu,\tau|M_i)$, $i\!<\!4$)
      are specified based on the posterior of a previous, separate
      meta-analysis, reflecting the corresponding information. This
      previous meta-analysis is again conducted using the vague
      prior~$p(\mu,\tau|M_4)$.

    \begin{figure*}[t!]
      \begin{center}
        \includegraphics[width=0.9\linewidth]{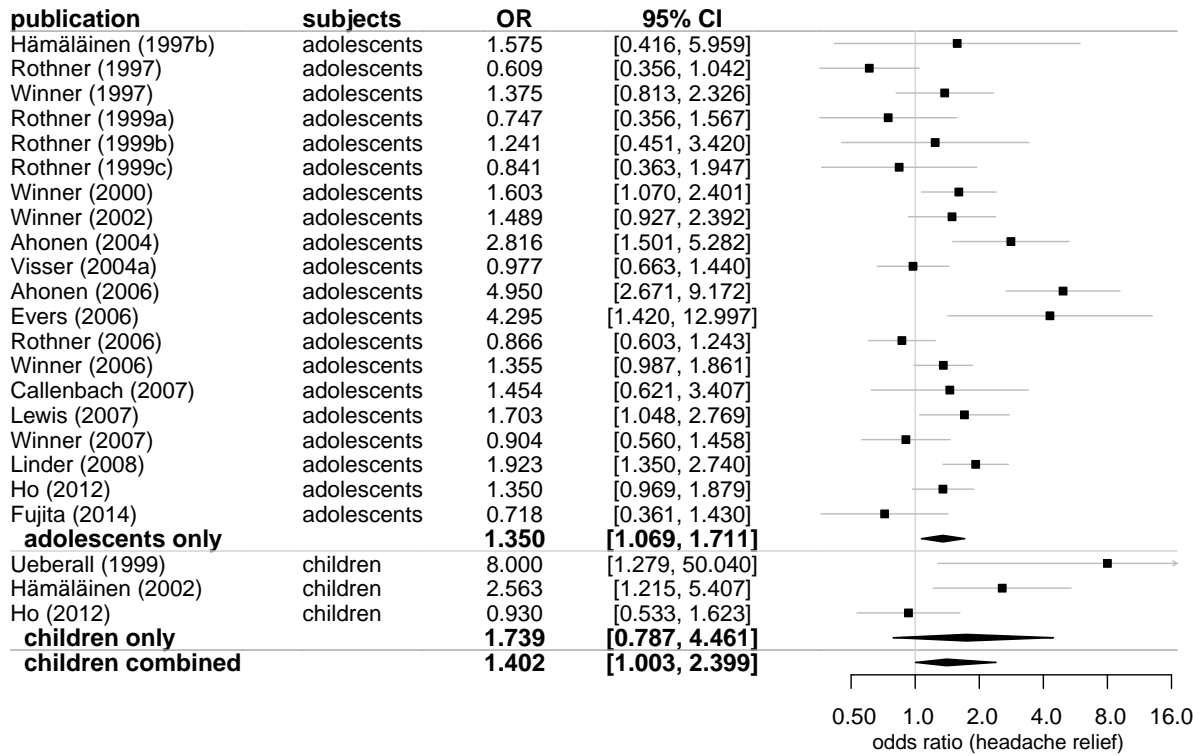}     
        \caption{\label{fig:RicherForest} Forest plot for the migraine
          example data introduced in Sec.~\ref{sec:richer}.  The last
          three studies (Ueberall (1999), H\"{a}m\"{a}l\"{a}inen
          (2002) and Ho (2012)) are the paediatric ones of primary
          interest, while the top 20~studies are the adolescents'
          studies constituting the external information that may
          potentially be extrapolated. The endpoint of interest here
          is the odds ratio of headache relief for triptans
          vs.\ placebo. The combined estimate here is based on the
          2-component model with $p(M_1)=0.5$ (see also row~VII in
          Fig.~\ref{fig:RicherVariations}).}
      \end{center}
    \end{figure*}

      Considering the relationship between \emph{source} and
      \emph{target} data sets, a range of scenarios is conceivable;
      here we concentrate on a few simple possibilities. Firstly, the
      two data sets may be completely unrelated, so that we have two
      pairs of parameters ($\mus$/$\taus$ and $\mut$/$\taut$), and by
      analyzing one data set we cannot learn anything about the other
      set's parameters. On the other hand, effect and heterogeneity
      may be identical in both populations ($\mut = \mus$ and $\taut =
      \taus$), so that we can pool the data (MAC approach), or
      equivalently, use the posterior from one analysis as the prior
      for the other (MAP approach). It may also be possible that only
      the effect or only the heterogeneity parameter are shared
      between the two populations. For the analysis of target data,
      this results in four possible models:
      \begin{itemize}
        \item $M_1$: $\mut = \mus$, $\taut = \taus$\\
              (informative prior for $\mut$ and $\taut$; ``complete pooling'')
        \item $M_2$: $\mut = \mus$, $\taut\neq\taus$\\ 
              (informative prior for $\mut$ only;
              ``effect pooling'')
        \item $M_3$: $\mut\neq\mus$, $\taut = \taus$\\ 
              (informative prior for $\taut$ only;
              ``heterogeneity pooling'')
        \item $M_4$: $\mut\neq\mus$, $\taut\neq\taus$\\ 
              (vague prior; ``standalone analyses'')
      \end{itemize}
      These four (conditional) parameter models are associated with
      prior probabilities $p(M_i)$.

      The eventual model setup then includes the specification of the
      vague prior component ($p(\mu,\tau|M_4)$), and of the prior
      probabilities for the four model components~$M_i$. Some of the
      model probabilities~$p(M_i)$ may be set to zero, especially for
      components~2 or~3.

      While the parameter models $M_1$ and $M_4$ may be very obvious,
      reflecting the commonly faced choice between data pooling and
      separate analyses, the other two models deserve some more
      consideration.  Model~$M_3$ supports the analysis of target data
      by only informing the heterogeneity prior based on the source
      data, an approach that is familiar from previous proposals
      \cite{TurnerEtAl2015,RhodesEtAl2015}. Model~$M_2$ is technically
      similar, but the scenario may be harder to motivate: a case in
      which the main effect is identical, but the heterogeneity is
      different may not be very realistic. Also, the relative
      similarity of models $M_1$ and $M_2$ as well as $M_3$ and $M_4$
      especially in the context of relatively few observations
      (small~$\ks$ and~$\kt$) may be an argument in favour of a
      sparser model not necessarily considering all four components.

  \subsection{Computation}
    Inference within the NNHM may technically be approached in
    different ways, for example using stochastic integration via MCMC
    methods. In the following we will utilize the
    \texttt{rjags}\cite{R:rjags} and \texttt{bayesmeta}
    \textsf{R}~packages \cite{bayesmeta,Roever2017}.  Use of the
    \texttt{bayesmeta} package simplifies computations, but it is only
    applicable for a subset of models (namely those with
    $p(M_2)\!=\!0$).

    \begin{figure*}
      \begin{center}
        \includegraphics[width=0.40\linewidth]{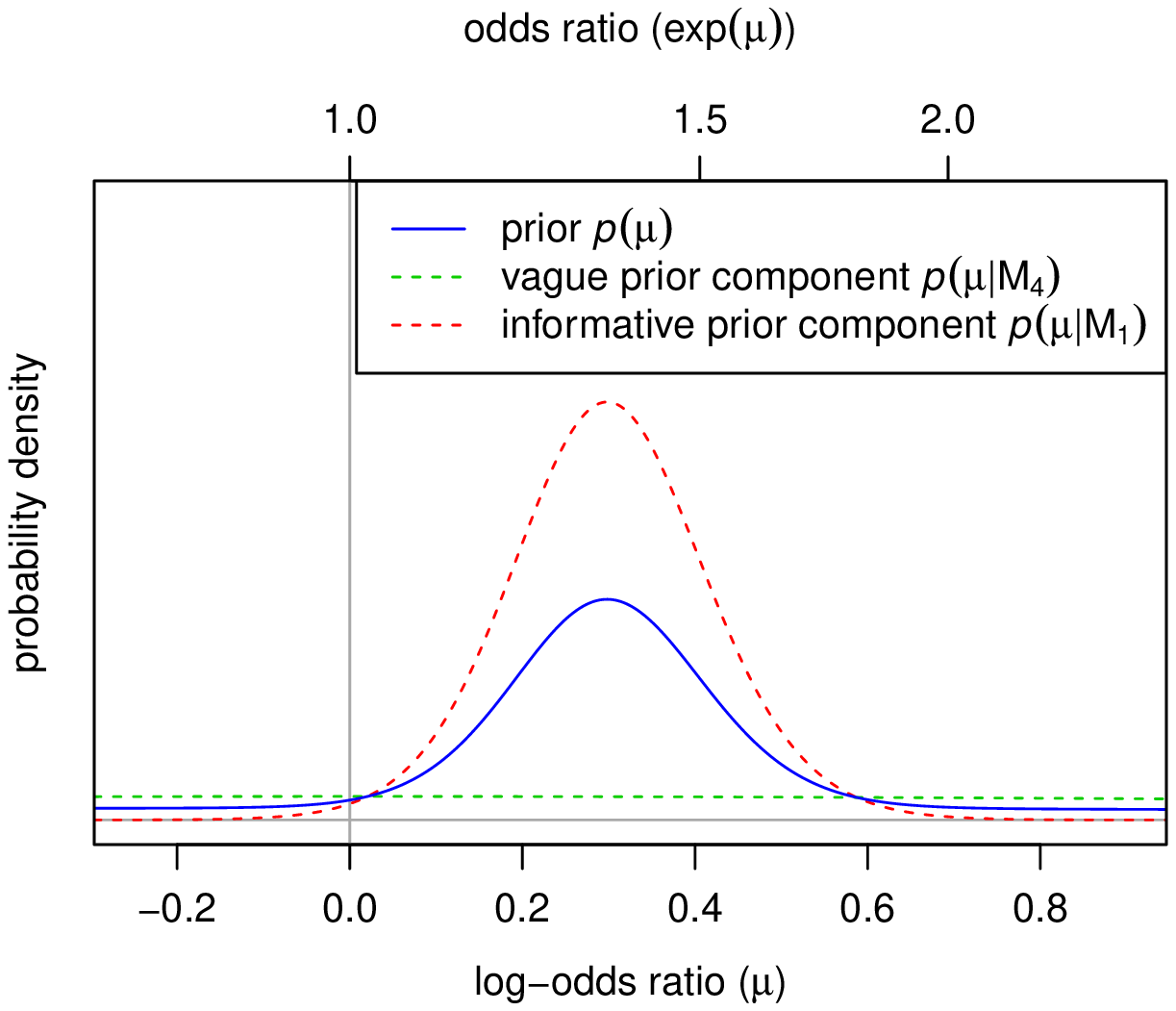}     
        \hspace{0.05\linewidth}
        \includegraphics[width=0.40\linewidth]{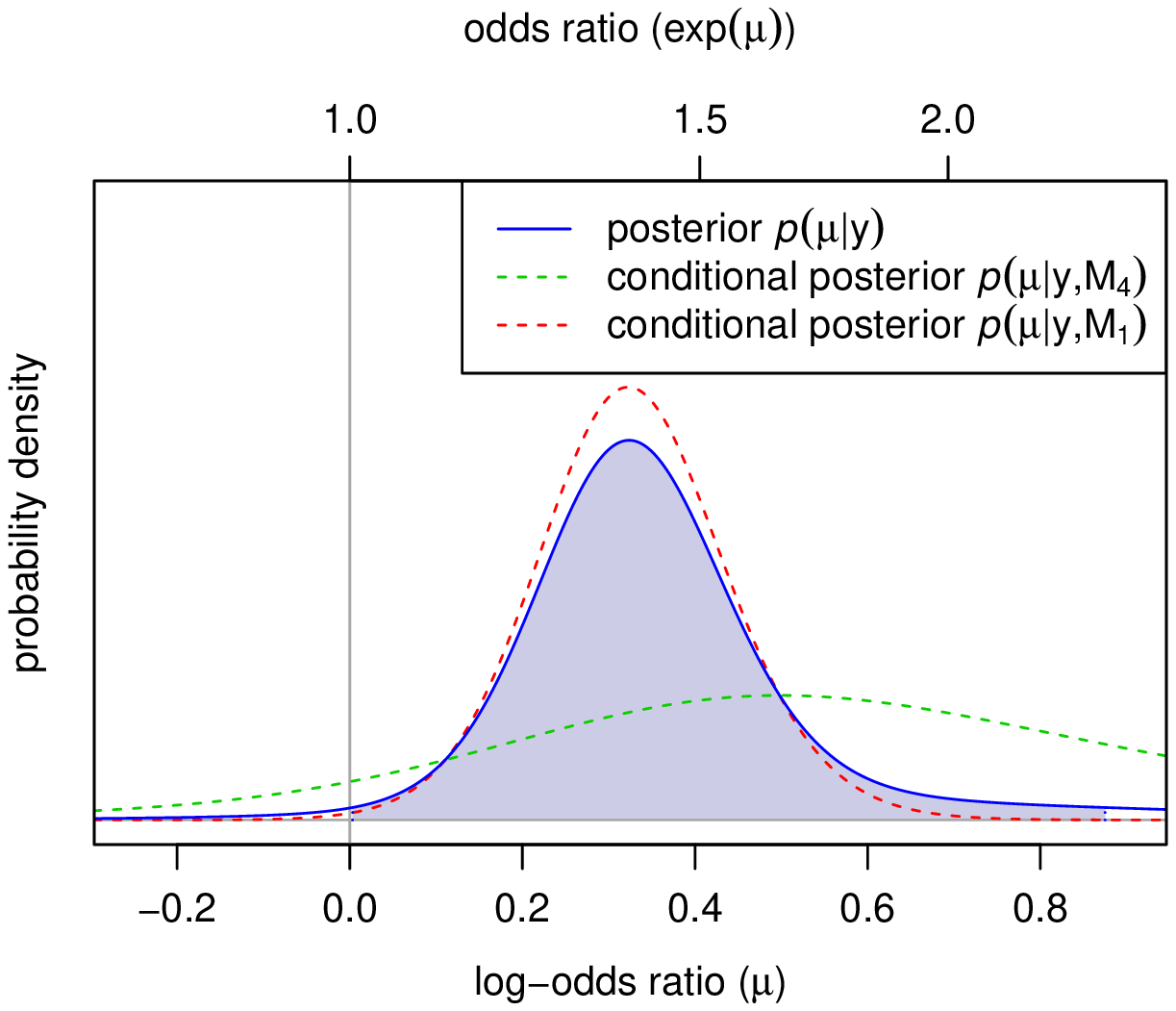} 
        \caption{\label{fig:RicherDensities} Prior (left) and
          posterior (right) densities for the example from
          Sec.~\ref{sec:richer}. The left panel shows the prior along
          with its two (informative and vague) mixture components. The
          right panel shows the posterior, which again is a mixture of
          two (conditional) posteriors. The shaded area indicates the
          posterior 95\% CI.}
      \end{center}
    \end{figure*}

\section{Paediatric migraine case study}\label{sec:richer}
  \subsection{Background}\label{sec:richerBackground}
    The data considered in the following are due to the systematic
    review by Richer \textit{et al.}\cite{RicherEtAl2016}, reporting
    on the evidence on the effect of pharmacological medications for
    the treatment of acute migraine attacks in children and
    adolescents.  Among the analyses performed is a comparison of the
    effect of triptans (vs.~placebo) on the proportion of patients
    reporting \emph{headache relief}. While 20~studies were found
    quoting odds ratios for the effect in adolescents ($12$--$17$
    years of age), only 3~studies reported on headache relief in
    children ($<\!12$~years of age).  The relevant data (numbers of
    cases and events, and the derived ORs) are reproduced in
    Tab.~\ref{tab:RicherData} in the appendix, the effect sizes are
    also illustrated in Fig.~\ref{fig:RicherForest}.

    Now suppose we are interested in estimating the effect in
    children.  Three studies constitute only a small data basis on
    which to judge efficacy
    \cite{FriedeRoeverWandelNeuenschwander2017a}, and indeed, a
    meta-analysis of the three studies fails to yield a conclusive
    result; 
    while the odds ratio is estimated at $1.739$ (a beneficial
    effect), the credible interval ([$0.787$, $4.461$]) is very wide and
    still includes a neutral effect of~$1.0$.  
    It would be desirable
    if the additional adolescent studies could possibly be used to
    clarify whether a clinically relevant effect is present or not.
    In order to summarize the evidence from the adolescent studies, we
    can perform a meta-analysis using the vague prior from
    Sec.~\ref{sec:vagueprior}, which yields an estimate and 95\% CI
    for the OR in adolescents of 1.350 [1.069, 1.711].

  \subsection{Analysis setup} \label{sec:prelim}
    In the present example there is a small number of target studies
    that are of primary interest (3~paediatric studies), and in
    addition there is another set of potentially relevant additional
    source studies available (20~studies performed in adolescent
    patients).  The additional data provide external information, but
    it is not evident \emph{a~priori} to what extent these are
    directly comparable and whether extrapolation is valid.  It may be
    plausible that the distinction between age groups is somewhat
    arbitrary, and that the effect~($\mu$) is indeed the same (or at
    least very similar) in both age groups.  This uncertainty is
    reflected in the analysis model setup via the specification of
    prior probabilities $p(M_i)$ for the different model components.

    The data are in the following analyzed using a mixture of two
    components, considering the cases of ``complete pooling'' ($M_1$),
    and of two ``standalone'' analyses ($M_4$). A~priori, we assume a
    probability of $p(M_1)=0.5$ for the joint model, and $p(M_4)=0.5$
    for unrelated effects in adolescents and adults.  With a
    substantial amount of scepticism associated with the prior
    information, we consider this a conservative choice
    \cite{NeuenschwanderEtAl2016a}.  We will also consider alternative
    prior setups later on.

    \begin{figure*}
      \begin{center}
        \includegraphics[width=0.40\linewidth]{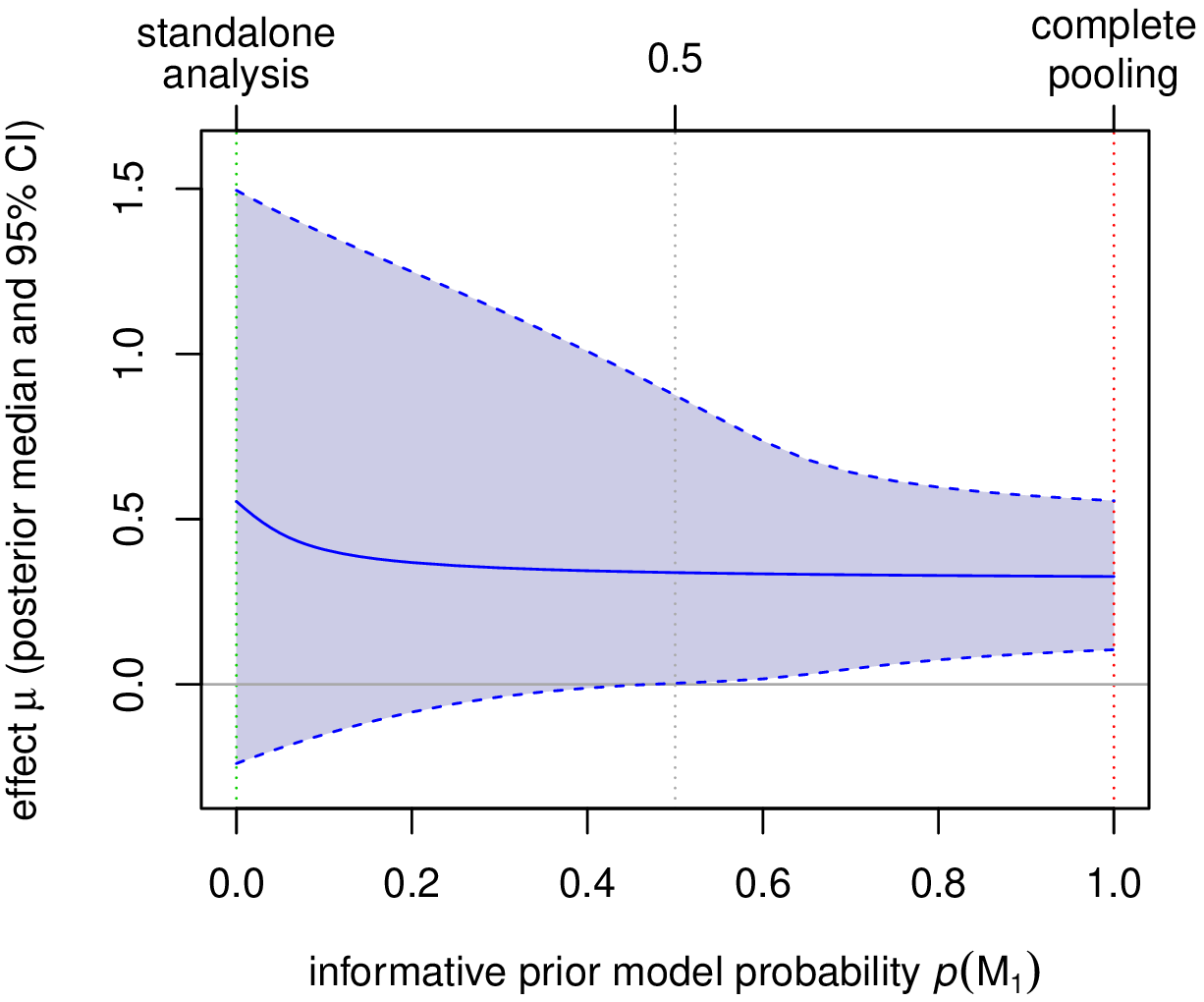}     
        \hspace{0.05\linewidth}
        \includegraphics[width=0.40\linewidth]{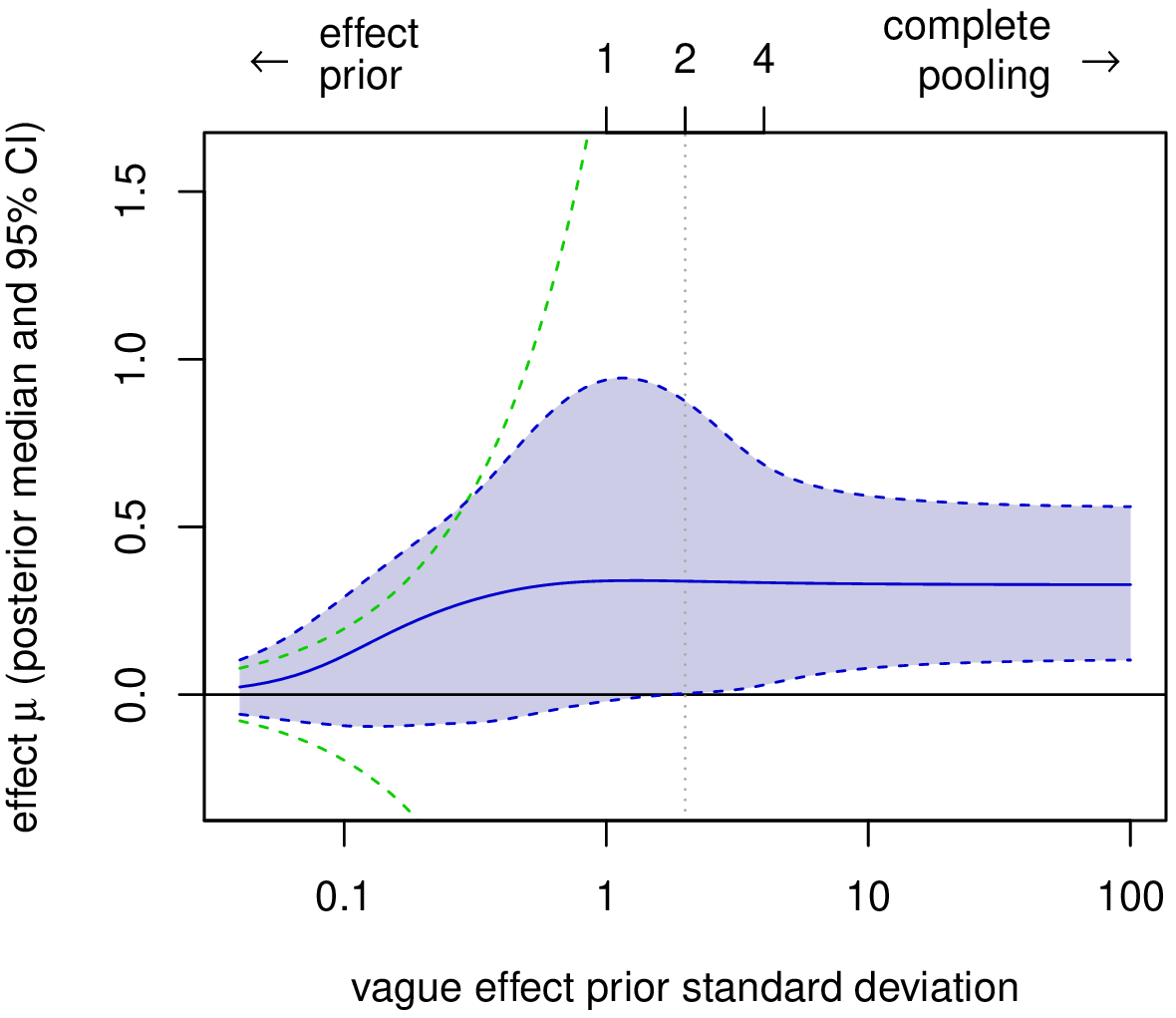}     
        \caption{\label{fig:RicherDiagnostics} Diagnostic plots
          showing the effect of varying the prior probability $p(M_1)$
          (left panel) and the vague prior standard deviation
          $\sqrt{\operatorname{Var}(\mu|M_4)}$ (right panel) in the
          migraine example from Sec.~\ref{sec:richer}.  The solid blue
          line shows the resulting effect estimate (log-OR), the
          shaded area indicates the corresponding 95\%~CI\@.  The
          dashed green lines indicate a prior 95\% range.}
      \end{center}
    \end{figure*}

    The two components as well as the resulting mixture prior for the
    effect ($\mu$) are also shown in Fig.~\ref{fig:RicherDensities}
    (left panel).  The vague component (according to
    Sec.~\ref{sec:vagueprior}) is normal with zero mean and standard
    deviation of~2. The informative component results as the posterior
    from the analysis of the adolescents' data (see
    Sec.~\ref{sec:richerBackground}).

  \subsection{Analysis results} \label{sec:richerResults}
    Analysis of the children's data under the two prior components
    yields a Bayes factor of~5.1 in favour of the ``pooling'' model
    ($M_1$). Given our prior specifications, this implies a posterior
    probability of $p(M_1|y)=0.837$ for the joint model. Technically,
    the effect's posterior distribution then results as a
    correspondingly weighted mixture of the two conditional posteriors
    $p(\mu|y, M_i)$; these are also shown in
    Fig.~\ref{fig:RicherDensities} (right panel), where one can see
    how the posterior mixture here turns out to be dominated by the
    ``informative'' conditional $p(\mu|y,M_1)$.  The estimated effect
    then is at an OR of $1.402$, with a 95\% credible interval of
    [$1.003$, $2.399$]. The estimate is also shown along with the data
    in Fig.~\ref{fig:RicherForest}.  The \textsf{R}~code to reproduce
    these results is available in the Appendix.

    \begin{figure*}[t]
      \begin{center}
        \includegraphics[width=0.95\linewidth]{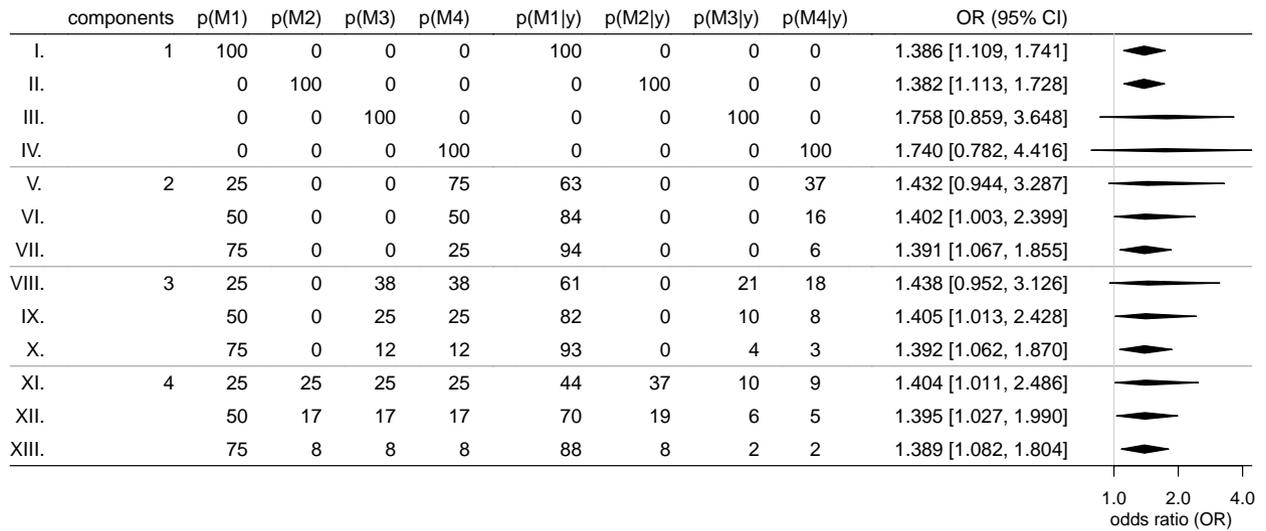}     
        \caption{\label{fig:RicherVariations} Analyses of the migraine
          data using different prior weights~$p(M_i)$ (in percent) for
          the model components
           ($M_1$ = ``complete pooling'', $M_2$ = ``effect-only pooling'', $M_3$ = ``heterogeneity-only pooling'', $M_4$ = ``standalone analyses''). 
          The resulting posterior model probabilities $p(M_i|y)$ as
          well as corresponding estimate and CI of the OR are shown.}
      \end{center}
    \end{figure*}

  \subsection{Sensitivity analyses} \label{sec:sensitivity}
    \subsubsection{Varying model probabilities}
    We may now also investigate the effect of changing the prior
    probability $p(M_1)$ expressing the a~priori expectations of
    whether and how the two data sources may be aggregated.
    Fig.~\ref{fig:RicherDiagnostics} (left panel) illustrates the
    resulting effect estimates (log-OR) and credible intervals when
    varying the prior probability~$p(M_1)$. In the extreme cases of
    $p(M_1)\!=\!1$ and $p(M_1)\!=\!0$, we are left with the simple
    models of ``complete pooling'' ($M_1$) and two separate
    ``standalone analyses'' ($M_4$), respectively. In between, we can
    see how the estimate is more or less ``shrunk'' towards the pooled
    estimate, and as long as $p(M_1)\!\geq\! 0.5$, the credible
    interval indicates a positive treatment effect.

    \subsubsection{Varying the vague prior specification}
    Besides varying the models' prior weight via~$p(M_1)$, we can also
    investigate the effect of different specifications of the effect's
    prior standard deviation (as defined in Sec.~\ref{sec:vagueprior})
    by varying it from its initial value of~$2$\@.
    Fig.~\ref{fig:RicherDiagnostics} (right panel) illustrates the
    effect on the log-OR estimate and credible interval. As expected,
    a very small standard deviation eventually shrinks the estimate
    towards the prior mean (left side). For very large variances, one
    can see the effect of Lindley's paradox (see
    Sec.~\ref{sec:mixprior}): the marginal likelihood of the
    paediatric data under $M_4$ decreases, and the ``pooling''
    component~$M_1$ dominates the posterior.

    The current setting of a standard deviation of~$2$ corresponds to
    a~priori 95\% probability roughly within a range of odds ratios
    between $\frac{1}{50}$ and~$50$. Varying the range of ORs by a
    factor of two ($25$ or~$100$ instead of~$50$) would correspond to
    prior standard deviations of~$1.64$ or~$2.35$,
    respectively. Doubling the standard deviation to a value of~$2$
    would correspond to an increase of the prior range up to values of
    $50^2\!=\!2500$ already, so the range of plausible values is
    probably not too far from~$2$.

    \subsubsection{Considering more than two components}
    One may consider more than two components in the model, in order
    to account for the possibility of a different association between
    adolescents' and children's data.  Fig.~\ref{fig:RicherVariations}
    presents the resulting estimates for a range of models with only a
    single or up to four components. The first columns show the prior
    and resulting posterior probabilities ($p(M_i)$ and $p(M_i|y)$)
    for the different model components. Row~VI shows the results
    discussed in Sec~\ref{sec:richerResults}, while rows~I, IV, V
    and~VII are among the estimates also shown in
    Fig.~\ref{fig:RicherDiagnostics} (left panel).

    One can see that the results based on parameter models~$M_1$
    and~$M_2$ as well as those for~$M_3$ and~$M_4$ (first four rows)
    are very similar. Based on the given data, it seems hard to
    distinguish between these pairs of models; when given equal prior
    probabilities, the posterior probabilities tend to be similar as
    well, as one can see e.g. in row~XI; using only components $M_1$
    and $M_4$ instead leads to a very similar estimate
    (row~VI). Again, as soon as $p(M_1)\!\geq\! 0.5$, all credible
    intervals indicate a positive effect estimate.

\section{Paediatric transplantation case study}\label{sec:crins}
  \subsection{Background}
\begin{figure*}[t]
    \begin{center}
      \includegraphics[width=0.9\linewidth]{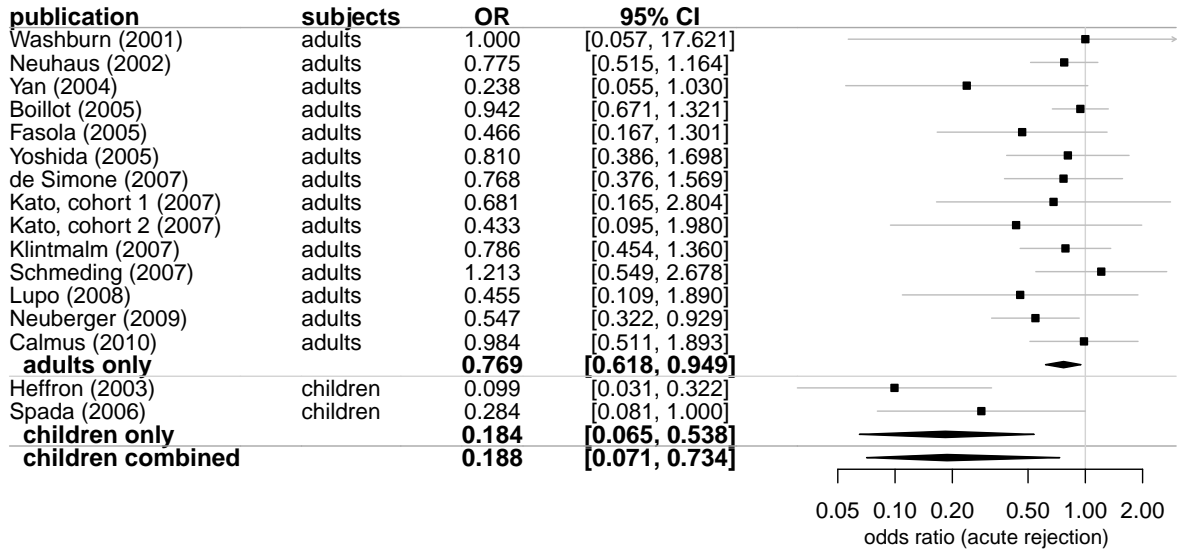}     
    \caption{\label{fig:CrinsForest} Forest plot for the
      transplantation data discussed in Sec.~\ref{sec:crins}.  The
      last two studies (Heffron (2003) and Spada (2006)) are the
      paediatric ones that are in the focus of the investigation,
      while the top 14~studies are based on adults and constitute the
      external information that may (potentially) be extrapolated. The
      endpoint of interest here is the number of acute rejection
      events. The combined estimate here is based on the 2-component
      model with $p(M_1)=0.5$.}
    \end{center}
  \end{figure*}

  The data considered in the following example are from Goralczyk
  \textit{et al.}\cite{GoralczykEtAl2011} and Crins \textit{et
    al.}\cite{CrinsEtAl2014}, and these illustrate a case of an
  apparent prior/data conflict.  Both studies were meta-analyses
  investigating the effect of Interleukin-2 receptor antagonists
  (IL-2RA) on the reported frequencies of acute rejection (AR)
  reactions after liver transplantation.  Both reviews included
  controlled studies; here we focus on the subset of randomized
  controlled trials.  The earlier publication \cite{GoralczykEtAl2011}
  was concerned with adults, while the more recent publication
  \cite{CrinsEtAl2014} was on paediatric patients.
  Tab.~\ref{tab:CrinsData} in the Appendix shows the relevant data
  (numbers of cases and events, and derived ORs) of 16 randomized
  studies; the effect sizes are also illustrated in
  Fig.~\ref{fig:CrinsForest}.  Considering the case of paediatric
  liver transplantation, there are only two studies available. Given
  the present body of evidence based on adult patients (14~studies),
  where the immune reaction may be similar, it is of interest to allow
  for this additional data to possibly inform the meta-analysis of the
  two paediatric studies.

  \begin{figure*}
    \begin{center}
      \includegraphics[width=0.40\linewidth]{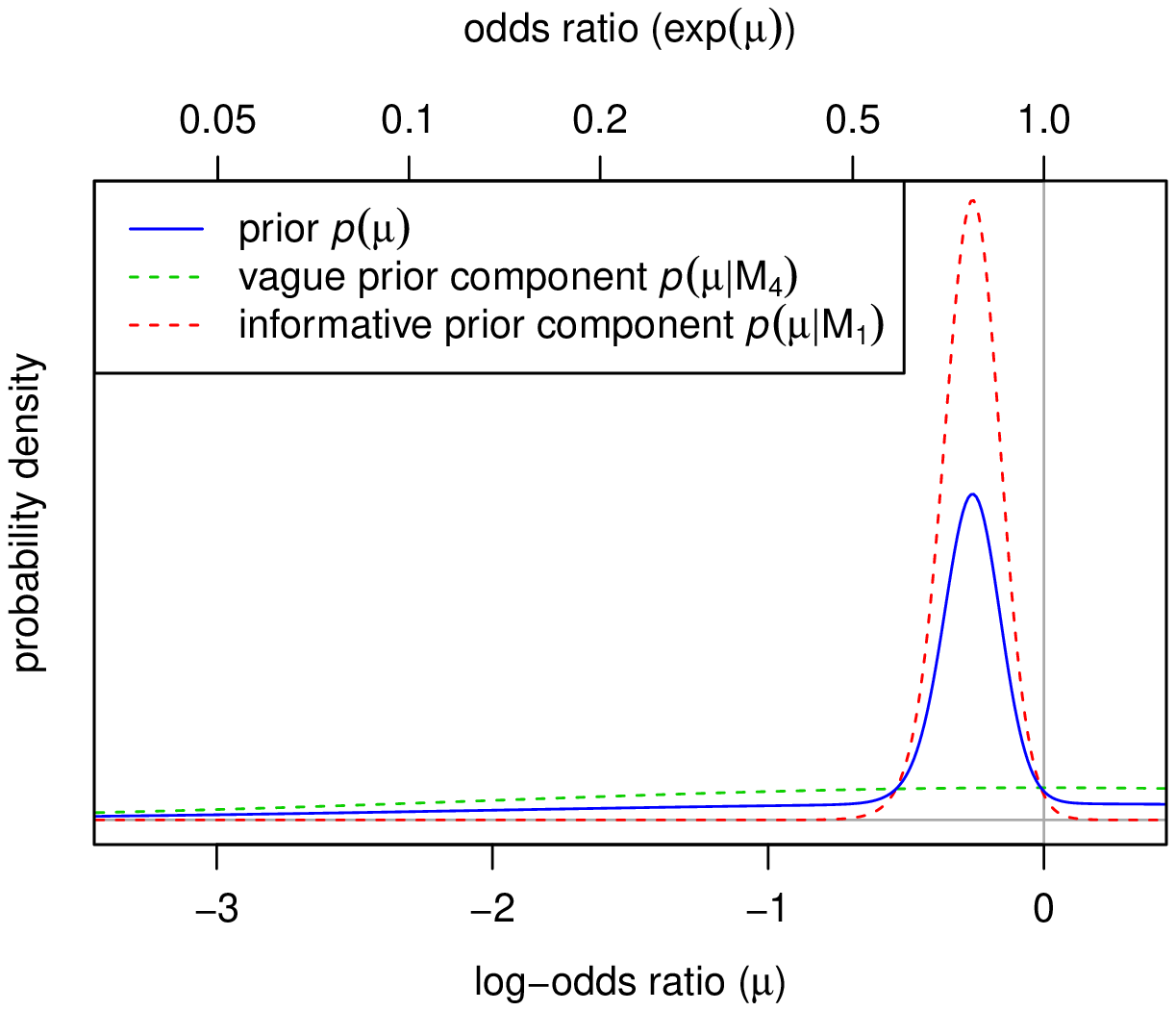}     
      \hspace{0.02\linewidth}
      \includegraphics[width=0.40\linewidth]{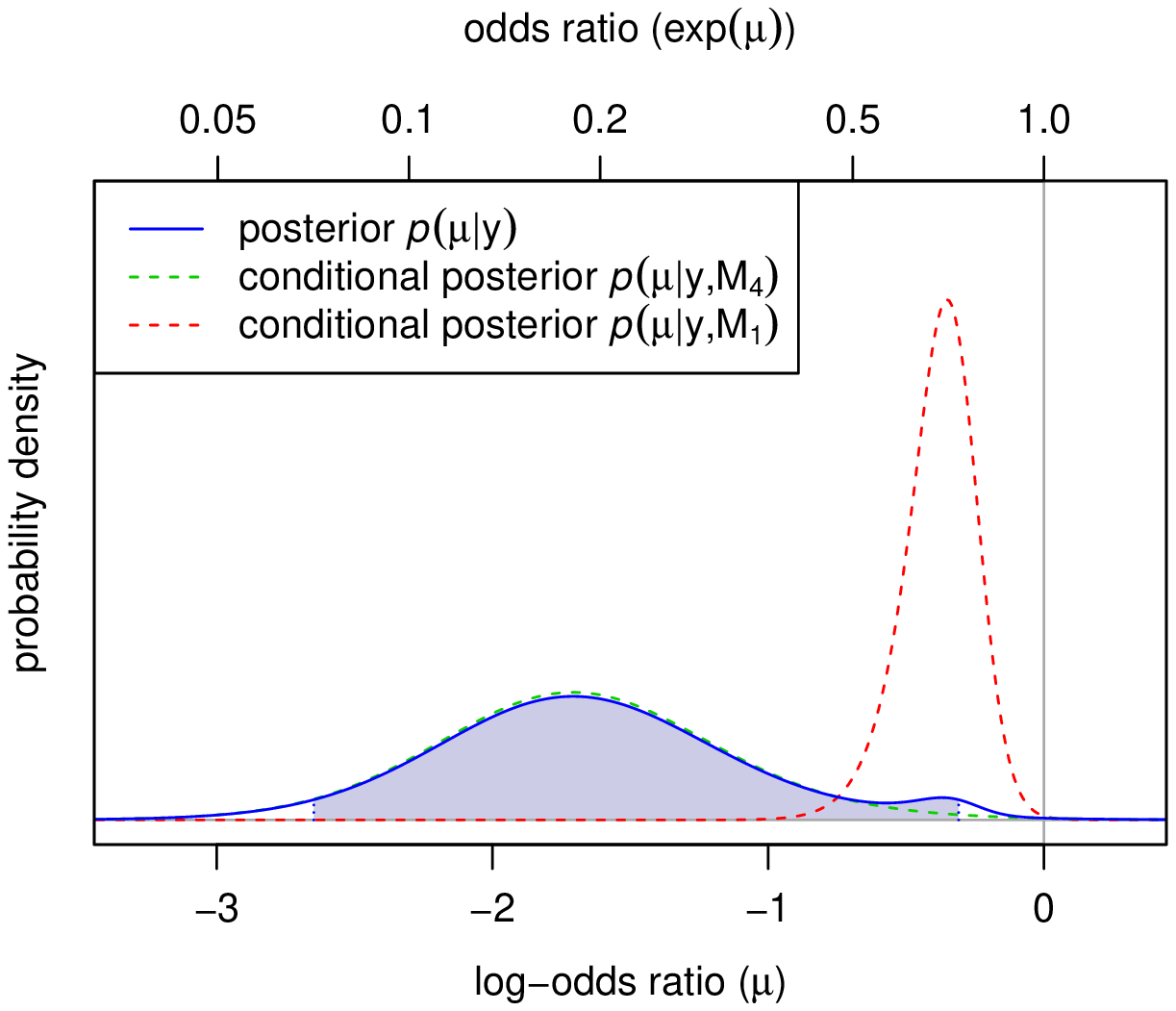} 
      \caption{\label{fig:CrinsDensities} Prior (left) and posterior
        (right) densities for the example from
        Sec.~\ref{sec:crins}. The left panel shows the prior along
        with its two (informative and vague) mixture components. The
        right panel shows the posterior, which again is a mixture of
        two (conditional) posteriors. The shaded area indicates the
        posterior 95\% CI.}
    \end{center}
  \end{figure*}

  \subsection{Analysis setup and results}
  We use an analogous setup as in the previous analysis and consider
  the two models~$M_1$ and~$M_4$. The prior used to analyze the
  paediatric data again is a heavy-tailed mixture of a vague component
  (as described in Sec.~\ref{sec:prelim}) and the posterior derived
  from the adult data. The weight of the informative prior component
  again is taken to be $p(M_1)=0.5$. From the adults' data alone we
  get posterior mean and standard deviation for~$\mu$ of \mbox{-0.266}
  and \mbox{0.109}, respectively; 
  the estimate and 95\% CI for the OR is \mbox{0.769} [0.618, 0.949].  
  The resulting prior as well as
  conditional and marginal posteriors are illustrated in
  Fig.~\ref{fig:CrinsDensities}.  The circumstances here differ from
  the previous example, as the paediatric data now look rather
  different from the external source data: we observe a larger effect
  (greater reduction of rejection reactions) in children than in
  adults (see Fig.~\ref{fig:CrinsForest}).

  \begin{figure}[t]
    \begin{center}
      \includegraphics[width=0.90\linewidth]{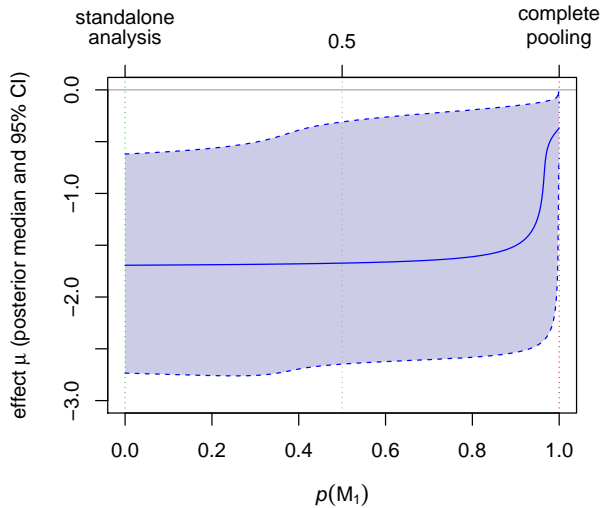}     
      \caption{\label{fig:CrinsSensitivity}Diagnostic plot showing
        the effect of varying the prior probability $p(M_1)$ 
        on the resulting effect estimate (posterior median and CI)
        in the
        transplantation example from Sec.~\ref{sec:crins}.}
    \end{center}
  \end{figure}

  Although a~priori we would tend to assume that we might also analyse
  the data jointly ($p(M_1)=0.5$), the data lead us to revise our view
  ($p(M_1|y)=0.031$); the Bayes factor here is at 30.9 in favour of
  the ``standalone analysis'' model ($M_4$).
  Fig.~\ref{fig:CrinsSensitivity} illustrates the effect of varying
  the prior certainty~$p(M_1)$: one can see that the ``pooling'' model
  ($M_1$) is heavily discounted unless one had a very strong prior in
  its favour (say, $p(M_1)>95\%$).  Including the possibility of $M_1$
  here mostly has the effect of widening the posterior to include the
  range suggested by the external data, and only an extremely strong
  prior confidence will actually shrink the posterior towards complete
  pooling. Effectively this leads to more cautious conclusions,
  including the possibility of a less pronouced effect as suggested by
  the external information.  Based on this analysis we get an estimate
  and 95\% CI for the OR in children of 0.188 [0.071, 0.734].

    \begin{table}
    \centering
    \caption{\label{tab:simulSetup} The parameter values used in the
      four simulation scenarios ($S_1$--$S_4$, corresponding to models
      $M_1$--$M_4$) discussed in Sec.~\ref{sec:simul}. The boldface
      figures indicate the parameters that differ between source and
      target.}
    \begin{tabular}{cccccc}
      \toprule
                 &                & \multicolumn{2}{c}{source}     & \multicolumn{2}{c}{target}\\[-0.5ex]
                 &                & \multicolumn{2}{c}{parameters} & \multicolumn{2}{c}{parameters}\\
      \cmidrule(lr){3-4}     \cmidrule(lr){5-6}
      scenario   & (model)        & $\mus$ & $\taus$ & $\mut$ & $\taut$ \\ \midrule
      $S_1$      & ($M_1$)     &    0.25 &      0.2 &   0.25 &      0.2 \\ 
      $S_2$      & ($M_2$)     &    0.25 &      0.2 &   0.25 &      \textbf{0.5} \\ 
      $S_3$      & ($M_3$)     &    0.25 &      0.2 & \textbf{1.0} &      0.2 \\ 
      $S_4$      & ($M_4$)     &    0.25 &      0.2 & \textbf{1.0} &      \textbf{0.5} \\ \bottomrule
    \end{tabular}
    \end{table}

  \begin{table*}[t]
  \centering \setlength{\tabcolsep}{4pt}
  \caption{\label{tab:simul} Coverage (\%), and CI width in the four simulation scenarios ($S_1$--$S_4$) and using a number of analysis settings (differing numbers of studies as well as differing prior settings; similar to Fig.~\ref{fig:RicherVariations}).}
  \begin{tabular}{lrrrrrrrrrrrrr}
    \toprule
    \# studies && \multicolumn{4}{c}{prior $p(M_i)$ (\%)} & \multicolumn{2}{c}{$S_1$} & \multicolumn{2}{c}{$S_2$} & \multicolumn{2}{c}{$S_3$} & \multicolumn{2}{c}{$S_4$} \\
    \cmidrule(lr){3-6} \cmidrule(lr){7-8} \cmidrule(lr){9-10} \cmidrule(lr){11-12} \cmidrule(lr){13-14}
     ($\ks+\kt$) && $M_1$ & $M_2$ & $M_3$ & $M_4$ & coverage & width & coverage & width & coverage & width & coverage & width \\ 
    \midrule
    $10+3$ & I.    & 100 & 0 & 0 & 0 & 97.1 & (0.66) & 96.0 & (0.69) & 10.8 & (0.73) & 15.6 & (0.77) \\ 
    & IV.   & 0 & 0 & 0 & 100 & 98.7 & (1.59) & 95.0 & (1.67) & 98.7 & (1.59) & 94.5 & (1.67) \\ \addlinespace[1.0ex]
    & V.    & 25 & 0 & 0 & 75 & 99.6 & (1.29) & 97.5 & (1.41) & 95.4 & (1.53) & 89.7 & (1.59) \\ 
    & VI.   & 50 & 0 & 0 & 50 & 99.5 & (1.06) & 97.9 & (1.19) & 89.5 & (1.45) & 81.9 & (1.50) \\ 
    & VII.  & 75 & 0 & 0 & 25 & 98.8 & (0.86) & 97.9 & (0.98) & 76.4 & (1.33) & 70.4 & (1.38) \\ \addlinespace[1.0ex]
    & VIII. & 25 & 0 & 38 & 38 & 99.3 & (1.25) & 96.8 & (1.35) & 94.8 & (1.47) & 88.1 & (1.52) \\ 
    & IX.   & 50 & 0 & 25 & 25 & 99.4 & (1.04) & 97.6 & (1.16) & 89.0 & (1.41) & 80.5 & (1.45) \\ 
    & X.    & 75 & 0 & 12 & 12 & 98.8 & (0.85) & 97.7 & (0.97) & 76.2 & (1.31) & 69.2 & (1.35) \\ \addlinespace[1.0ex]
    & XI.   & 25 & 25 & 25 & 25 & 99.5 & (1.05) & 98.3 & (1.15) & 89.3 & (1.44) & 80.5 & (1.48) \\ 
    & XII.  & 50 & 17 & 17 & 17 & 99.0 & (0.92) & 98.0 & (1.03) & 81.7 & (1.37) & 73.7 & (1.40) \\ 
    & XIII. & 75 & 8 & 8 & 8 & 98.5 & (0.79) & 97.6 & (0.89) & 68.6 & (1.26) & 63.3 & (1.30) \\ 
    \midrule
    $3+3$ & I.    & 100 & 0 & 0 & 0 & 98.1 & (1.06) & 96.2 & (1.12) & 77.5 & (1.16) & 74.9 & (1.22) \\ 
    & IV.   & 0 & 0 & 0 & 100 & 98.7 & (1.59) & 95.0 & (1.67) & 98.7 & (1.59) & 94.5 & (1.67) \\ \addlinespace[1.0ex]
    & V.    & 25 & 0 & 0 & 75 & 99.1 & (1.42) & 96.6 & (1.51) & 97.3 & (1.54) & 92.8 & (1.61) \\ 
    & VI.   & 50 & 0 & 0 & 50 & 99.1 & (1.29) & 96.8 & (1.38) & 95.2 & (1.49) & 90.0 & (1.56) \\ 
    & VII.  & 75 & 0 & 0 & 25 & 98.7 & (1.18) & 96.7 & (1.27) & 91.2 & (1.42) & 86.2 & (1.48) \\ \addlinespace[1.0ex]
    & VIII. & 25 & 0 & 38 & 38 & 99.0 & (1.40) & 96.2 & (1.49) & 97.1 & (1.52) & 92.4 & (1.59) \\ 
    & IX.   & 50 & 0 & 25 & 25 & 99.0 & (1.28) & 96.7 & (1.37) & 95.0 & (1.48) & 89.8 & (1.54) \\ 
    & X.    & 75 & 0 & 12 & 12 & 98.7 & (1.18) & 96.7 & (1.26) & 91.2 & (1.41) & 86.0 & (1.47) \\ \addlinespace[1.0ex]
    & XI.   & 25 & 25 & 25 & 25 & 99.1 & (1.28) & 96.9 & (1.36) & 95.0 & (1.49) & 89.7 & (1.55) \\ 
    & XII.  & 50 & 17 & 17 & 17 & 98.9 & (1.21) & 96.8 & (1.30) & 92.6 & (1.44) & 87.3 & (1.50) \\ 
    & XIII. & 75 & 8 & 8 & 8 & 98.6 & (1.14) & 96.6 & (1.22) & 89.1 & (1.38) & 84.1 & (1.44) \\ 
    \bottomrule
  \end{tabular}
  \end{table*}

\section{Simulation study}\label{sec:simul}
  \subsection{Simulation setup}
    In order to gain more insight into the behaviour of the
    extra\-po\-la\-tion model, we run simulations reflecting the four
    parameter models considered. We use three or ten
    ($\ks\in\{3,10\}$) source studies with $\mus\!=\!0.25$ and
    $\taus\!=\!0.2$.  Of primary interest are three ($\kt\!=\!3$)
    target studies with equal or different parameters $\mut$ and
    $\taut$ as shown in Tab.~\ref{tab:simulSetup}.  Estimates~$y_i$
    are generated (according to the model, see Sec.~\ref{sec:remodel})
    on a continuous scale, and standard errors ($s_i$) are drawn
    uniformly between~0.2 and~1.0; for binary (log-OR) outcomes, this
    roughly corresponds to sample sizes between~16 and~400. We then
    investigate resulting coverages and mean widths of 95\% CIs based
    on $10\,000$ replications each for each scenario, yielding a
    simulation error of 0.22 percentage points for the coverages.

    In order to check validity of our computations, we also generated
    data sets with parameters drawn from the prior distribution and
    checked for proper coverage of the resulting credible intervals,
    which should be exact by construction
    \citep{Dawid1982,CookGelmanRubin2006}. Data were generated based
    on parameters drawn from the vague prior distribution, either
    independently or identically for target and source data with
    probability~$0.5$, corresponding to the setting used e.g. in
    Sec.~\ref{sec:prelim}.

  \subsection{Simulation results}
    CI coverages and widths for $\ks\!=\!10$ source studies
    are shown in Tab.~\ref{tab:simul}. 
    As expected, ``na\"{i}ve'' extrapolation based only on model~$M_1$
    fails especially in scenarios $S_3$ and $S_4$ (row~I), while a
    standalone analysis of the target data only yields proper
    coverage, but at the cost of much wider CIs (model~$M_4$, row~IV).
    Combining several prior components to a mixture then allows to
    gain in CI width, while coverage probability is slightly reduced
    in case of a prior/data (source/target) conflict.
    Again, we know that on average over the corresponding prior
    distribution, the coverage will be at exactly~95\% by
    construction.
    Coverage is above the nominal~95\% in scenario~$S_1$ as well as
    the very similar~$S_2$, and it is lower in scenarios~$S_3$
    and~$S_4$.
    As already apparent in the example application in
    Sec.~\ref{sec:richer}, due to the limited amount of data considered,
    models~$M_1$ and~$M_2$ as well as models~$M_3$ and~$M_4$ are
    barely distinguishable. Tab.~\ref{tab:prob} (see Appendix) shows
    that (e.g.\ in row~XI), when the pairs of models have equal prior
    probabilities, the resulting posterior probabilities tend to be
    very similar as well, i.e., the corresponding Bayes factors are
    close to unity. The model similarity is also evident in the
    resulting estimates when comparing e.g. rows~VI, IX and~XI in
    Tab.~\ref{tab:simul} or Fig.~\ref{fig:RicherVariations}.

    In addition to the simulations with $\ks\!=\!10$ ``source''
    studies, we also investigated the performance for a smaller
    external evidence base of only $\ks=3$ studies.  The behaviour
    with respect to CI coverage and length is qualitatively similar,
    but not quite as pronounced. Regarding the posterior
    probabilities~$p(M_i|y)$ (Tab.~\ref{tab:prob} in the Appendix) it
    is apparent that these are mostly determined by the prior
    probabilities and less affected by the data. Based on the little
    data only, the models apparently are hardly distinguishable, and
    consequently extra care should be taken to specify priors
    reasonably.

    In the $10\,000$ simulations with parameter values drawn from the
    2-component mixture prior, 94.96\% of credible intervals covered
    the true parameter values in the setting of $\ks\!+\!\kt=10+3$
    studies, and coverage was at 95.08\% for $\ks\!+\!\kt=3+3$
    studies, which in both cases is within the range expected for a
    nominal 95\% coverage.
    This was expected by construction of the CI, as mentioned above.

\section{Discussion}

  Mixture priors provide a means to support an analysis using external
  information in a robust manner \cite{SchmidliEtAl2014}. We have
  showcased in two examples how the approach may easily be implemented
  in evidence synthesis by utilizing the simplicity of the mixture
  model, which implies that the posterior again constitutes a
  \emph{model average}, a weighted mixture of the conditional
  posteriors based on the prior components. In the meta-analysis
  context, this means that off-the-shelf software may be used to
  perform the main computations, which then only need to be
  re-combined.  MCMC methods only become necessary when a 4-component
  model is desired.  The resulting procedure provides a transparent
  and robust data-driven approach to analysis that has the potential
  to either boost or discount relevant prior information, depending on
  its apparent compatibility.

  The examples discussed here were intentionally restricted to simple
  random-effects meta-analysis with log-odds ratio endpoints. The same
  approach is readily extended to other types of endpoints that are
  conventionally analyzed using a random-effects model.  The main
  point here was primarily to demonstrate the approach of using
  heavy-tailed \cite{OHaganPericchi2012} or mixture priors
  \cite{SchmidliEtAl2014}, its simplicity and its potential.  More
  generally, the approach demonstrated here shows a way of overcoming
  the \texttt{bayesmeta} package's restriction to normal effect
  priors, which is due to the semi-analytic implementation
  \cite{Roever2017,RoeverFriede2017}. Robust priors for the
  heterogeneity parameter may already be implemented by simply using
  e.g. heavy-tailed half-Student-\mbox{$t$}, half-Cauchy, or Lomax
  distributions.

  When looking for example at the meta-analyses published in the
  \emph{Cochrane library}, a large number of investigations include
  additional analyses of pre-defined subgroups of studies in addition
  to an overall estimate \citep{TurnerEtAl2015}. Such cases are
  examples in which there may be a benefit from borrowing of
  information on effect or heterogeneity. The obvious danger here is
  that, being presented with subgroup as well as overall estimates,
  the practitioner may effectively perform the extrapolation in a
  rather intransparent manner based on eyeballing data and estimates.

  While the use of two prior components may often be reasonable and
  sufficient, the setup has the disadvantage that prior/data conflict
  is confounded for effect and heterogeneity. Extending to more
  general formulations including more components may provide some more
  flexibility, if desired. However, the simulations showed that slight
  model variations may not be distinguishable or may not make a
  noticeable difference if data are sparse.
  Based on the principle of parsimony (Occam's razor), one may then
  want to give preference to simpler, sparser model formulations.
  
  In a Bayesian analysis, proper coverage of credible intervals is, by
  construction, guaranteed \emph{conditional on the prior
    distribution}; this implies that the long-run coverage will be
  exact if the data-generating parameter values are repeatedly drawn
  from the prior distribution\citep{CookGelmanRubin2006}.  However,
  coverage probability is not necessarily at the nominal level when
  data are generated repeatedly based on single constant parameter
  values (which is the common frequentist
  requirement\citep{NeymanPearson1933}).  By constructing the prior as
  informative, and complementing the informative prior with a
  ``robustifying'' vague component, we expect the coverage to exceed
  the nominal credible level when conditioning on the informative
  component only, and to be lower conditional on the vague component.
  As usual, inferences will be reasonable and consistent when model
  and prior are specified sensibly; in the present case it is
  especially crucial to also specify the vague prior component
  realistically. In contrast to immediate intuition, a larger prior
  variance is not necessarily a more conservative choice, due to
  \emph{Lindley's paradox}. In order to enhance robustness, the prior
  probability of the vague component should be increased instead.

  Many other variations of the approach are conceivable. The external
  information does not necessarily need to come from a second
  meta-analysis, but could also be based on other types of
  data. Likewise, the ``main'' analysis does not need to be a
  meta-analysis, but could also be a single study with a meta-analysis
  informing the prior, which would lead to an approach very similar to
  the original setup discussed by Schmidli \textit{et
    al.}\cite{SchmidliEtAl2014}.

  The presented approach however is no substitute for a careful check
  of appropriateness of possible data pooling.  Note that in the
  transplantation example a joint analysis may already be highly
  questionable on theoretical grounds. While in the preceding migraine
  example, it is conceivable that with adjusted dosing a comparable
  effect may be achieved in adolescents and children, in the
  transplantation context, indications and surgical practice differ
  between adults and children in a range of aspects, so that a
  similarity of effect may already be doubtful \emph{a~priori}.  Even
  if the different data themselves may not be obviously contradicting,
  pooling always also requires a theoretical justification, and
  plausibility should be reflected in the model setup (here especially
  in the weighting of prior components).

\section*{Acknowledgments}
  The authors wish to thank Beat Neuenschwander and Heinz Schmidli for
  helpful comments.  This research has received funding from the EU's
  7th Framework Programme for research, technological development and
  demonstration under grant agreement number FP HEALTH 2013-602144
  with project title (acronym) ``Innovative methodology for small
  populations research'' (InSPiRe).

\subsection*{Author contributions}
  CR, SW and TF conceived the concept of this study, CR conducted all
  numerical evaluations for the examples and the simulations, and
  drafted the manuscript. TF and SW critically reviewed and made
  substantial contributions to the manuscript. All authors commented
  on and approved the final manuscript.

\subsection*{Financial disclosure}
  FP HEALTH 2013-602144
  ``Innovative methodology for small populations research'' (InSPiRe).

\subsection*{Conflict of interest}
  Tim Friede and Christian R\"{o}ver declare no conflict of interest. 
  Simon Wandel is employed by Novartis Pharma AG, and owns stocks thereof.

\clearpage

\appendix

\section{Mixture posterior derivation\label{sec:mixtureAppendix}}
  Consider a setup where the prior distribution is a two-component
  mixture
  \begin{equation}
    p(\vartheta) \;=\; p(\vartheta|M_a)\, p(M_a) + p(\vartheta|M_b)\, p(M_b)
  \end{equation}
  and interest lies in determining the posterior $p(\vartheta|y)$
  based on data~$y$ and some likelihood function $p(y|\vartheta)$.
  The parameters' posterior distribution then is given by
    \begin{eqnarray}
      && p(\vartheta|y) \;=\; \frac{p(y|\vartheta)\; p(\vartheta)}{\int p(y|\vartheta)\,p(\vartheta) \, \differential \vartheta} \\
      &=& \frac{p(y|\vartheta)\; \bigl(p(\vartheta|M_a)\, p(M_a) + p(\vartheta|M_b)\, p(M_b)\bigr)}{\int p(y|\vartheta)\, \bigl(p(\vartheta|M_a)\, p(M_a) + p(\vartheta|M_b)\, p(M_b)\bigr)\, \differential \vartheta}\\
      &=& \frac{p(M_a)\, p(y|\vartheta)\,p(\vartheta|M_a)+ p(M_b)\, p(y|\vartheta)\,p(\vartheta|M_b)}{p(M_a)\, \int p(y|\vartheta)\,p(\vartheta|M_a)\, \differential \vartheta + p(M_b)\,\int p(y|\vartheta) \,p(\vartheta|M_b)\, \differential \vartheta}\\
      &=& \frac{p(M_a)\, p(y|\vartheta)\,p(\vartheta|M_a)+ p(M_b)\, p(y|\vartheta)\,p(\vartheta|M_b)}{p(M_a)\, p(y|M_a) + p(M_b)\, p(y|M_b)}\\
      &=& \frac{p(M_a)\, p(y|\vartheta)\,p(\vartheta|M_a)}{p(M_a)\, p(y|M_a) + p(M_b)\,p(y|M_b)} \\ 
      &&    + \frac{p(M_b)\, p(y|\vartheta)\,p(\vartheta|M_b)}{p(M_a)\, p(y|M_a) + p(M_b)\,p(y|M_b)}\\
      &=& \frac{p(y|\vartheta)\; p(\vartheta|M_a)}{p(y|M_a)} \; \frac{p(M_a)\, p(y|M_a)}{p(M_a)\, p(y|M_a) + p(M_b)\,p(y|M_b)} \nonumber \\
       && +\frac{p(y|\vartheta)\; p(\vartheta|M_b)}{p(y|M_b)} \; \frac{p(M_b)\,p(y|M_b)}{p(M_a)\, p(y|M_a) + p(M_b)\,p(y|M_b)} \\
      &=& p(\vartheta|y, M_a) \; p(M_a|y)  +  p(\vartheta|y, M_b) \; p(M_b|y) \mbox{,}
    \end{eqnarray}
  which again is a mixture of the two conditional posterior
  distributions $p(\vartheta|y, M_a)$ and $p(\vartheta|y, M_b)$.

\section{Example data\label{sec:dataAppendix}}
  Tab.~\ref{tab:RicherData} shows the paediatric migraine example data
  due to Richer \textit{et al.}\cite{RicherEtAl2016}, which are
  discussed in Sec.~\ref{sec:richer}.
  The paediatric transplantation example data due to Crins \textit{et
    al.}\cite{CrinsEtAl2014}, which are used in Sec.~\ref{sec:crins},
  are shown in Tab.~\ref{tab:CrinsData}.

  \begin{center}
  \begin{table*}[!h]
  \caption{\label{tab:RicherData}Data set due to Richer
    \textit{et~al.}\cite{RicherEtAl2016}; the numbers of events and total
    numbers of patients in treatment and control groups are shown
    along with the corresponding derived logarithmic odds ratios
    (log-ORs) and confidence intervals.}  \centering
  \begin{tabular}{lcrrc}
   \toprule
              & \multicolumn{1}{c}{patient} & \multicolumn{1}{c}{triptan} & \multicolumn{1}{c}{placebo} & log-OR \\ \addlinespace[-0.3ex]
  publication & \multicolumn{1}{c}{type}    & \multicolumn{1}{c}{group}     & \multicolumn{1}{c}{group}   & (95\% CI)\\ 
   \midrule
  H\"{a}m\"{a}l\"{a}inen (1997b) & adolescents &   7 /  23 &   5 /  23 &  0.454 [-0.876, 1.785] \\ 
  Rothner (1997)                 & adolescents & 113 / 226 &  46 /  74 & -0.496 [-1.034, 0.041] \\ 
  Winner (1997)                  & adolescents & 111 / 222 &  32 /  76 &  0.318 [-0.207, 0.844] \\ 
  Rothner (1999a)                & adolescents &  96 / 186 &  20 /  34 & -0.292 [-1.033, 0.449] \\ 
  Rothner (1999b)                & adolescents &  17 /  62 &   7 /  30 &  0.216 [-0.797, 1.230] \\ \addlinespace[1.0ex] 
  Rothner (1999c)                & adolescents &  23 /  66 &  14 /  36 & -0.174 [-1.014, 0.666] \\ 
  Winner (2000)                  & adolescents & 243 / 377 &  69 / 130 &  0.472 [0.068, 0.876] \\ 
  Winner (2002)                  & adolescents &  98 / 149 &  80 / 142 &  0.398 [-0.076, 0.872] \\ 
  Ahonen (2004)                  & adolescents &  53 /  83 &  32 /  83 &  1.035 [0.406, 1.664] \\ 
  Visser (2004a)                 & adolescents & 159 / 233 & 165 / 240 & -0.024 [-0.412, 0.364] \\ \addlinespace[1.0ex]
  Ahonen (2006)                  & adolescents &  71 /  96 &  35 /  96 &  1.599 [0.982, 2.216] \\ 
  Evers (2006)                   & adolescents &  18 /  29 &   8 /  29 &  1.458 [0.350, 2.565] \\ 
  Rothner (2006)                 & adolescents & 262 / 480 &  93 / 160 & -0.144 [-0.506, 0.218] \\ 
  Winner (2006)                  & adolescents & 316 / 483 & 141 / 242 &  0.304 [-0.013, 0.621] \\ 
  Callenbach (2007)              & adolescents &  19 /  46 &  15 /  46 &  0.375 [-0.477, 1.226] \\ \addlinespace[1.0ex]
  Lewis (2007)                   & adolescents &  97 / 148 &  67 / 127 &  0.533 [0.046, 1.019] \\ 
  Winner (2007)                  & adolescents &  82 / 144 &  79 / 133 & -0.101 [-0.579, 0.377] \\ 
  Linder (2008)                  & adolescents & 383 / 544 &  94 / 170 &  0.654 [0.300, 1.008] \\ 
  Ho (2012)                      & adolescents & 167 / 284 & 147 / 286 &  0.300 [-0.031, 0.631] \\ 
  Fujita (2014)                  & adolescents &  23 /  74 &  27 /  70 & -0.331 [-1.019, 0.357] \\ \addlinespace[1.0ex]
  Ueberall (1999)                & children    &  12 /  14 &   6 /  14 &  2.079 [0.246, 3.913] \\ 
  H\"{a}m\"{a}l\"{a}inen (2002)  & children    &  38 /  59 &  24 /  58 &  0.941 [0.195, 1.688] \\ 
  Ho (2012)                      & children    &  53 /  98 &  57 / 102 & -0.073 [-0.630, 0.485] \\ 
   \bottomrule
  \end{tabular}
  \end{table*}

  \begin{table*}[!h]
  \caption{\label{tab:CrinsData}Data set due to Goralczyk \textit{et
      al.}\cite{GoralczykEtAl2011} and Crins \textit{et
      al.}\cite{CrinsEtAl2014}. The numbers of events and total
    numbers of patients in treatment and control groups are shown
    along with the corresponding derived odds ratios (ORs) and
    confidence intervals.}
  \begin{tabular}{lcrrc} 
  \toprule
              & \multicolumn{1}{c}{patient} & \multicolumn{1}{c}{IL-2RA} & \multicolumn{1}{c}{control} & odds ratio \\ \addlinespace[-0.3ex]
  publication & \multicolumn{1}{c}{type}    & \multicolumn{1}{c}{group}     & \multicolumn{1}{c}{group}   & (95\% CI)\\ 
  \midrule
  Washburn (2001)       & adults   &  1 /  15 &  1 /  15 & 1.000 [0.057, 17.62] \\ 
  Neuhaus (2002)        & adults   & 74 / 188 & 88 / 193 & 0.775 [0.515, 1.164] \\ 
  Yan (2004)            & adults   &  3 /  24 &  9 /  24 & 0.238 [0.055, 1.030] \\ 
  Boillot (2005)        & adults   & 89 / 351 & 92 / 347 & 0.942 [0.671, 1.321] \\ 
  Fasola (2005)         & adults   & 13 /  46 & 11 /  24 & 0.466 [0.167, 1.301] \\ \addlinespace[1.0ex]
  Yoshida (2005)        & adults   & 17 /  72 & 21 /  76 & 0.810 [0.386, 1.698] \\ 
  de Simone (2007)      & adults   & 17 /  95 & 21 /  95 & 0.768 [0.376, 1.569] \\ 
  Kato, cohort 1 (2007) & adults   &  7 /  15 &  9 /  16 & 0.681 [0.165, 2.804] \\ 
  Kato, cohort 2 (2007) & adults   &  3 /  16 &  8 /  23 & 0.433 [0.095, 1.980] \\ 
  Klintmalm (2007)      & adults   & 80 / 153 & 46 /  79 & 0.786 [0.454, 1.360] \\ \addlinespace[1.0ex]
  Schmeding (2007)      & adults   & 29 /  51 & 25 /  48 & 1.213 [0.549, 2.678] \\ 
  Lupo (2008)           & adults   &  4 /  26 &  6 /  21 & 0.455 [0.109, 1.890] \\ 
  Neuberger (2009)      & adults   & 28 / 168 & 45 / 168 & 0.547 [0.322, 0.929] \\ 
  Calmus (2010)         & adults   & 23 /  98 & 24 / 101 & 0.984 [0.511, 1.893] \\ \addlinespace[1.0ex]
  Heffron (2003)        & children & 14 /  61 & 15 /  20 & 0.099 [0.031, 0.322] \\ 
  Spada (2006)          & children &  4 /  36 & 11 /  36 & 0.284 [0.081, 1.000] \\ 
   \bottomrule
  \end{tabular}
  \end{table*}
  \end{center}

\section{Additional simulation results: model proabilities}
  Tab.~\ref{tab:prob} below shows the average model probabilities
  corresponding to the simulation results shown in
  Tab.~\ref{tab:simul} (Sec.~\ref{sec:simul}).

  \begin{table*}[ht] \footnotesize
  \centering \setlength{\tabcolsep}{4pt}
  \caption{\label{tab:prob} Average posterior model probabilities $p(M_i|y)$ (in \%) in the four simulation scenarios ($S_1$--$S_4$).}
  \begin{tabular}{lrrrrrrrrrrrrrrrrrrrrr}
    \toprule
    \# studies & & \multicolumn{4}{c}{prior $p(M_i)$ (\%)} & \multicolumn{4}{c}{$S_1$} & \multicolumn{4}{c}{$S_2$} & \multicolumn{4}{c}{$S_3$} & \multicolumn{4}{c}{$S_4$} \\
    \cmidrule(lr){3-6} \cmidrule(lr){7-10} \cmidrule(lr){11-14} \cmidrule(lr){15-18} \cmidrule(lr){19-22}
    ($\ks+\kt$) & & $M_1$ & $M_2$  & $M_3$  & $M_4$  & $\quad M_1$ & $M_2$  & $M_3$  & $M_4$  & $\quad M_1$ & $M_2$  & $M_3$  & $M_4$  & $\quad M_1$ & $M_2$  & $M_3$  & $M_4$  & $\quad M_1$ & $M_2$  & $M_3$  & $M_4$  \\ 
    \midrule
    $10+3$ & V.    & 25 &  0 &  0 & 75 & 55 &  0 &  0 & 45 & 48 &  0 &  0 & 52 & 27 &  0 &  0 & 73 & 27 &  0 &  0 & 73 \\ 
    & VI.   & 50 &  0 &  0 & 50 & 77 &  0 &  0 & 23 & 70 &  0 &  0 & 30 & 46 &  0 &  0 & 54 & 45 &  0 &  0 & 55 \\ 
    & VII.  & 75 &  0 &  0 & 25 & 90 &  0 &  0 & 10 & 86 &  0 &  0 & 14 & 65 &  0 &  0 & 35 & 62 &  0 &  0 & 38 \\ \addlinespace[1.0ex]
    & VIII. & 25 &  0 & 38 & 38 & 54 &  0 & 23 & 23 & 48 &  0 & 25 & 27 & 27 &  0 & 38 & 36 & 27 &  0 & 36 & 37 \\ 
    & IX.   & 50 &  0 & 25 & 25 & 76 &  0 & 12 & 12 & 70 &  0 & 14 & 15 & 46 &  0 & 28 & 27 & 45 &  0 & 27 & 28 \\ 
    & X.    & 75 &  0 & 12 & 12 & 90 &  0 &  5 &  5 & 86 &  0 &  7 &  7 & 64 &  0 & 18 & 17 & 62 &  0 & 18 & 19 \\ \addlinespace[1.0ex]
    & XI.   & 25 & 25 & 25 & 25 & 39 & 37 & 12 & 12 & 35 & 36 & 14 & 14 & 23 & 24 & 27 & 26 & 22 & 25 & 26 & 26 \\ 
    & XII.  & 50 & 17 & 17 & 17 & 65 & 20 &  7 &  7 & 60 & 21 &  9 &  9 & 43 & 16 & 21 & 20 & 41 & 17 & 21 & 21 \\ 
    & XIII. & 75 &  8 &  8 &  8 & 85 &  9 &  3 &  3 & 81 & 10 &  5 &  5 & 64 &  8 & 14 & 13 & 62 &  9 & 15 & 15 \\ 
    \midrule
    $3+3$ & V.    & 25 & 0 & 0 & 75 & 48 & 0 & 0 & 52 & 45 & 0 & 0 & 55 & 32 & 0 & 0 & 68 & 31 & 0 & 0 & 69 \\ 
    & VI.   & 50 & 0 & 0 & 50 & 72 & 0 & 0 & 28 & 69 & 0 & 0 & 31 & 53 & 0 & 0 & 47 & 52 & 0 & 0 & 48 \\ 
    & VII.  & 75 & 0 & 0 & 25 & 88 & 0 & 0 & 12 & 86 & 0 & 0 & 14 & 73 & 0 & 0 & 27 & 71 & 0 & 0 & 29 \\ \addlinespace[1.0ex]
    & VIII. & 25 & 0 & 38 & 38 & 48 & 0 & 26 & 26 & 45 & 0 & 27 & 28 & 31 & 0 & 35 & 34 & 31 & 0 & 34 & 35 \\ 
    & IX.   & 50 & 0 & 25 & 25 & 72 & 0 & 14 & 14 & 69 & 0 & 15 & 16 & 53 & 0 & 24 & 23 & 52 & 0 & 24 & 24 \\ 
    & X.    & 75 & 0 & 12 & 12 & 88 & 0 & 6 & 6 & 86 & 0 & 7 & 7 & 72 & 0 & 14 & 14 & 71 & 0 & 14 & 14 \\ \addlinespace[1.0ex]
    & XI.   & 25 & 25 & 25 & 25 & 37 & 35 & 14 & 14 & 35 & 35 & 15 & 15 & 26 & 27 & 23 & 23 & 26 & 27 & 23 & 24 \\ 
    & XII.  & 50 & 17 & 17 & 17 & 63 & 20 & 8 & 8 & 60 & 20 & 10 & 10 & 49 & 17 & 17 & 17 & 48 & 17 & 17 & 17 \\ 
    & XIII. & 75 & 8 & 8 & 8 & 83 & 9 & 4 & 4 & 81 & 9 & 5 & 5 & 71 & 8 & 10 & 10 & 70 & 8 & 11 & 11 \\ 
    \bottomrule
  \end{tabular}
  \end{table*}

\section{Example \textsf{R} code\label{sec:codeAppendix}}
  \subsection{Two-component mixture}
  The following \textsf{R}~code allows to reproduce the analysis from
  Sec.~\ref{sec:richer}. In addition to~\textsf{R}
  \cite{R-Manual}, the \texttt{metafor} \cite{Viechtbauer2010} and
  \texttt{bayesmeta} \cite{bayesmeta,Roever2017} packages are required. The other
  analyses are done completely analogously.
  { \footnotesize 
\begin{verbatim}
# read data:
RicherEtAl2016 <- read.csv("RicherEtAl2016.csv",
                           colClasses=c("study"="character"))
# compute effect sizes (log-ORs):
require("metafor")
effsize <- escalc(measure="OR",
                  ai=triptan.events, n1i=triptan.total,
                  ci=placebo.events, n2i=placebo.total,
                  slab=study, data=RicherEtAl2016)
# specify subset indices:
aidx <- (effsize[,"patients"]=="adolescents")
cidx <- (effsize[,"patients"]=="children")


########################
# main MA computations:
require("bayesmeta")

# standard deviation of vague prior:
vaguepriorsd <- 2

# meta analysis for adolescents only:
bma.adol <- bayesmeta(effsize[aidx,],
              mu.prior.mean=0, mu.prior.sd=vaguepriorsd,
              tau.prior=function(t){dhalfnormal(t,scale=0.5)})

# meta analysis for children only:
bma.child <- bayesmeta(effsize[cidx,],
               mu.prior.mean=0, mu.prior.sd=vaguepriorsd,
               tau.prior=function(t){dhalfnormal(t,scale=0.5)})

# joint meta analysis for all patients:
bma.joint <- bayesmeta(effsize,
               mu.prior.mean=0, mu.prior.sd=vaguepriorsd,
               tau.prior=function(t){dhalfnormal(t,scale=0.5)})

# assemble marginal likelihoods:
marginals <- c("M1: pooled"   = bma.joint$marginal,
               "M4: separate" = bma.adol$marginal
                                * bma.child$marginal)
print(marginals)
barplot(marginals, ylab="marginal likelihood")

# Bayes factor:
bf <- marginals[1] / marginals[2]
print(c("M1: pooled"=unname(bf), "M4: separate"=1/unname(bf)))

# specify prior (p(M1)):
prior.prob <- 0.50
prior.odds <- prior.prob / (1-prior.prob)

# determine posterior:
post.odds <- prior.odds * bf
post.prob <- post.odds / (post.odds+1)
print(post.prob)

# show prior & posterior probabilities:
print(matrix(c(prior.prob, 1-prior.prob,
               post.prob, 1-post.prob), 
             nrow=2, ncol=2, byrow=TRUE,
             dimnames=list(c("prior","posterior"),
                           c("M1: pooled","M4: separate"))))


##############################################################
# functions to compute cumulative distribution function etc.:

# cumulative distribution function (CDF):
cdf <- function(mu=0, prob = post.prob)
{
  dens <- function(x)
  {
    return(prob*bma.joint$dposterior(mu=x) 
           + (1-prob)*bma.child$dposterior(mu=x))
  }
  integral <- integrate(dens, lower=-Inf, upper=mu)
  return(integral$value)
}

# quantile function (inverse CDF):
invcdf <- function(p, xrange=c(-4,4), prob=post.prob)
{
  if ((cdf(xrange[1], prob=prob)<p)&(cdf(xrange[2], prob=prob)>p))
    result <- uniroot(function(x){cdf(x, prob=prob)-p},
                      lower=xrange[1], upper=xrange[2])$root
  else
    result <- NA
  return(result)
}

# function to determine shortest credible interval:
shortest.interval <- function(level=0.95,
                              min.p=0.001, prob=post.prob)
{
  intwidth <- function(left)
  {
    pleft <- cdf(left, prob=prob)
    right <- invcdf(level+pleft, prob=prob)
    return(right-left)
  }
  opti <- optimize(intwidth, lower=invcdf(min.p, prob=prob),
                   upper=invcdf(1-level, prob=prob))$minimum
  result <- c(opti, invcdf(level+cdf(opti, prob=prob), prob=prob))
  return(result)
}


########################################################
# compute eventual estimates (median and 95% interval):

# median & credible interval:
esti <- c("median" = invcdf(0.5, prob=post.prob),
          "lower" = NA, "upper" = NA)
# shortest interval:
esti[2:3] <- shortest.interval(prob=post.prob)
print(esti)
print(exp(esti))
\end{verbatim} }

  \subsection{Three-component mixture}
  The following \textsf{R}~code allows to reproduce the analysis
  shown in row~IX. of Fig.~\ref{fig:RicherVariations}, using a
  3-component mixture for the prior.
  { \footnotesize 
\begin{verbatim}
# read data:
RicherEtAl2016 <- read.csv("RicherEtAl2016.csv",
                           colClasses=c("study"="character"))
# compute effect sizes (log-ORs):
require("metafor")
effsize <- escalc(measure="OR",
                  ai=triptan.events, n1i=triptan.total,
                  ci=placebo.events, n2i=placebo.total,
                  slab=study, data=RicherEtAl2016)
# subset indices:
aidx <- (effsize[,"patients"]=="adolescents")
cidx <- (effsize[,"patients"]=="children")

########################
# main MA computations:
require("bayesmeta")

# standard deviation of vague prior:
vaguepriorsd <- 2

# meta analysis for adolescents only:
bma.adol <- bayesmeta(effsize[aidx,],
              mu.prior.mean=0, mu.prior.sd=vaguepriorsd,
              tau.prior=function(t){dhalfnormal(t,scale=0.5)})

# meta analysis for children only:
bma.child <- bayesmeta(effsize[cidx,],
               mu.prior.mean=0, mu.prior.sd=vaguepriorsd,
               tau.prior=function(t){dhalfnormal(t,scale=0.5)})

# joint meta analysis (BOTH parameters) for all patients:
bma.joint <- bayesmeta(effsize,
               mu.prior.mean=0, mu.prior.sd=vaguepriorsd,
               tau.prior=function(t){dhalfnormal(t,scale=0.5)})

# meta analysis for children only,
# heterogeneity prior from adolescents:
bma.child.t <- bayesmeta(effsize[cidx,],
                mu.prior.mean=0, mu.prior.sd=vaguepriorsd,
                tau.prior=function(t){bma.adol$dposterior(tau=t)})


# show some results
# (adolescents only / pooled / pooling tau only / children):
rbind("adolescents"=bma.adol$summary[,"mu"],
      "M1: pooled"=bma.joint$summary[,"mu"],
      "M3: pooled (tau only)"=bma.child.t$summary[,"mu"],
      "M4: separate"=bma.child$summary[,"mu"])


# assemble marginal likelihoods:
marginals <- c("M1: pooled (mu+tau)"   = bma.joint$marginal,
               "M3: pooled (tau only)" = bma.adol$marginal
                                         * bma.child.t$marginal,
               "M4: separate"          = bma.adol$marginal
                                         * bma.child$marginal)
print(marginals)
barplot(marginals, ylab="marginal likelihood")

# specify prior:
prior.prob <- c("M1: pooled (mu+tau)"   = 1/2,
                "M3: pooled (tau only)" = 1/4,
                "M4: separate"          = 1/4)

# determine posterior:
post.prob <- prior.prob * marginals
post.prob <- post.prob / sum(post.prob)

# show prior & posterior probabilities:
print(round(rbind("prior"=prior.prob, "posterior"=post.prob),3))


##############################################################
# functions to compute cumulative distribution function etc.:

# cumulative distribution function (CDF):
cdf <- function(mu=0, prob = post.prob)
{
  stopifnot(length(prob)==3, all(is.finite(prob)),
            all(prob>=0), all(prob<=1), sum(prob)==1)
  dens <- function(x)
  {
    return(prob[1]*bma.joint$dposterior(mu=x) 
           + prob[2]*bma.child.t$dposterior(mu=x)
           + prob[3]*bma.child$dposterior(mu=x))
  }
  integral <- integrate(dens, lower=-Inf, upper=mu)
  return(integral$value)
}

# quantile function (inverse CDF):
invcdf <- function(p, xrange=c(-4,4), prob=post.prob)
{
  if ((cdf(xrange[1], prob=prob)<p)&(cdf(xrange[2], prob=prob)>p))
    result <- uniroot(function(x){cdf(x, prob=prob)-p},
                      lower=xrange[1], upper=xrange[2])$root
  else
    result <- NA
  return(result)
}

# function to determine shortest credible interval:
shortest.interval <- function(level=0.95,
                              min.p=0.001, prob=post.prob)
{
  intwidth <- function(left)
  {
    pleft <- cdf(left, prob=prob)
    right <- invcdf(level+pleft, prob=prob)
    return(right-left)
  }
  opti <- optimize(intwidth, lower=invcdf(min.p, prob=prob),
                   upper=invcdf(1-level, prob=prob))$minimum
  result <- c(opti, invcdf(level+cdf(opti, prob=prob), prob=prob))
  return(result)
}


########################################################
# compute eventual estimates (median and 95% interval):

# median & symmetric interval:
esti <- c("median"  = invcdf(0.5, prob=post.prob),
          "lower" = NA, "upper" = NA)
# shortest interval (on log scale):
esti[2:3] <- shortest.interval(prob=post.prob)
print(esti)
print(exp(esti))
\end{verbatim} }

  \bibliographystyle{wileyNJD-AMA}
  \bibliography{../../literature/literature}

\end{document}